\documentclass[pre,twocolumn,amssymb]{revtex4}

\usepackage{mathtools}
\usepackage{amsmath}
\usepackage{float}
\usepackage{graphicx}
\usepackage{color}
\usepackage{epsfig}
\usepackage{latexsym}
\usepackage{bm}
\usepackage{ulem}
\usepackage{hyperref}
\usepackage{hyperref,graphicx}    
\usepackage{color}

\begin{document}

\newcommand{\rar}{\rightarrow}
\newcommand{\lar}{\leftarrow}
\newcommand{\rlh}{\rightleftharpoons}
\newcommand{\eref}[1]{Eq.~(\ref{#1})}%
\newcommand{\Eref}[1]{Equation~(\ref{#1})}%
\newcommand{\fref}[1]{Fig.~\ref{#1}} %
\newcommand{\Fref}[1]{Figure~\ref{#1}}%
\newcommand{\sref}[1]{Sec.~\ref{#1}}%
\newcommand{\Sref}[1]{Section~\ref{#1}}%
\newcommand{\aref}[1]{Appendix~\ref{#1}}%

\renewcommand{\ni}{{\noindent}}
\newcommand{\dprime}{{\prime\prime}}
\newcommand{\be}{\begin{equation}}
\newcommand{\ee}{\end{equation}}
\newcommand{\bea}{\begin{eqnarray}}
\newcommand{\eea}{\end{eqnarray}}
\newcommand{\nn}{\nonumber}
\newcommand{\bk}{{\bf k}}
\newcommand{\bQ}{{\bf Q}}
\newcommand{\q}{{\bf q}}
\newcommand{\s}{{\bf s}}
\newcommand{\bN}{{\bf \nabla}}
\newcommand{\bA}{{\bf A}}
\newcommand{\bE}{{\bf E}}
\newcommand{\bj}{{\bf j}}
\newcommand{\bJ}{{\bf J}}
\newcommand{\bs}{{\bf v}_s}
\newcommand{\bn}{{\bf v}_n}
\newcommand{\bv}{{\bf v}}
\newcommand{\la}{\left\langle}
\newcommand{\ra}{\right\rangle}
\newcommand{\dg}{\dagger}
\newcommand{\br}{{\bf{r}}}
\newcommand{\brp}{{\bf{r}^\prime}}
\newcommand{\bq}{{\bf{q}}}
\newcommand{\hx}{\hat{\bf x}}
\newcommand{\hy}{\hat{\bf y}}
\newcommand{\bS}{{\bf S}}
\newcommand{\cU}{{\cal U}}
\newcommand{\cD}{{\cal D}}
\newcommand{\bR}{{\bf R}}
\newcommand{\pll}{\parallel}
\newcommand{\sumr}{\sum_{\vr}}
\newcommand{\cP}{{\cal P}}
\newcommand{\cQ}{{\cal Q}}
\newcommand{\cS}{{\cal S}}
\newcommand{\ua}{\uparrow}
\newcommand{\da}{\downarrow}
\newcommand{\red}{\textcolor {red}}
\newcommand{\blu}{\textcolor {blue}}
\newcommand{\1}{{\oldstylenums{1}}}
\newcommand{\2}{{\oldstylenums{2}}}
\newcommand{\mDelta}{\varepsilon}
\newcommand{\m}{\tilde m}
\def\lsim {\protect \raisebox{-0.75ex}[-1.5ex]{$\;\stackrel{<}{\sim}\;$}}
\def\gsim {\protect \raisebox{-0.75ex}[-1.5ex]{$\;\stackrel{>}{\sim}\;$}}
\def\lsimeq {\protect \raisebox{-0.75ex}[-1.5ex]{$\;\stackrel{<}{\simeq}\;$}}
\def\gsimeq {\protect \raisebox{-0.75ex}[-1.5ex]{$\;\stackrel{>}{\simeq}\;$}}

\title{Density relaxation in conserved Manna sandpiles}

\author{Dhiraj Tapader$^{1}$} 
\author{Punyabrata Pradhan$^{1}$}
\email{punyabrata.pradhan@bose.res.in}
\author{Deepak Dhar$^{2}$}

\affiliation{$^1$Department of Theoretical Sciences, S. N. Bose National Centre for Basic Sciences, Block-JD, Sector-III, Salt Lake, Kolkata 700106, India \\ $^2$Department of Physics, Indian Institute of Science Education and Research, Pune, \\
Dr. Homi Bhabha Road, Pashan, Pune, 411008, India }

\begin{abstract}

\noindent{ We study relaxation of long-wavelength density perturbations in one dimensional conserved Manna sandpile. 
Far from criticality where correlation length $\xi$ is finite, relaxation of density profiles having wave numbers $k \rightarrow 0$ is diffusive, with relaxation time $\tau_R \sim k^{-2}/D$ with $D$ being the density-dependent bulk-diffusion coefficient.
Near criticality with $k \xi \gsim 1$, the bulk diffusivity diverges and the transport becomes anomalous; accordingly, the relaxation time varies as $\tau_R \sim k^{-z}$, with the dynamical exponent $z=2-(1-\beta)/\nu_{\perp} < 2$, where $\beta$ is the critical order-parameter exponent and and $\nu_{\perp}$ is the critical correlation-length exponent.
Relaxation of initially localized density profiles on infinite critical background exhibits a self-similar structure. In this case, the asymptotic scaling form of the time-dependent density profile is analytically calculated: we find that, at long times $t$, the width $\sigma$ of the density perturbation grows anomalously, i.e., $\sigma \sim t^{w}$, with the growth exponent $\omega=1/(1+\beta) > 1/2$. 
In all cases, theoretical predictions are in reasonably good agreement with simulations.
}

\typeout{polish abstract}

\end{abstract}

\maketitle

\section{Introduction}

Sandpile models \cite{Bak_PRL1987} were proposed to explain the ubiquitous power-law correlations in slowly evolving natural structures, such as mountain ranges \cite{Mountain}, river networks \cite{river}, and in the low-frequency `$1/f$' noise \cite{metal} and related dynamical phenomena \cite{Earthquake, Rain, Brain, Sethna_Nat2001, Peters_Nature2006, book-Jensen, Aschwanden-2013, Watkins-2016}. They are threshold-activated spatially extended discrete dynamical systems with lattice sites having grains or particles, which diffuse in the bulk through cascades of toppling events, called avalanches.
In the original version \cite{Bak_PRL1987, Dhar_PRL1990, Dhar_PhysicaA1999}, the systems are driven by slow addition of grains, which get dissipated at the boundaries. However, in the conserved or {\it fixed energy} version \cite{Dickman_PRL1998}, though the microscopic dynamics in the bulk remains the same, there is {\it no} dissipation and the total mass (number of grains) remains conserved in the system. 
In this paper, we consider a stochastic variant of conserved-mass sandpiles - the celebrated conserved Manna sandpile \cite{Manna_JPhysA1991, Dickman_PRE2001, Dickman_PRE2002}, which constitutes a paradigm for nonequilibrium systems undergoing an absorbing phase transition \cite{Marro_Dickman, Dickman_PRE1998}. That is, upon decreasing the global density below a critical value, the system goes from a dynamically {\it active} steady state to an {\it absorbing} state, devoid of any activity.

The conserved Manna sandpile has generated considerable interest in the past, especially concerning the questions of the universality class and the  formulation of the corresponding field-theoretic description of the system \cite{ Kardar_PRL1989, Kardar_PRA1992, Dickman_PRL1998, C-DP_Munoz2008, Ledoussal_PRL2015}. 
In fact, there are many universality classes  in different sandpile models, depending on details of toppling rules. While the questions concerning universality in sandpiles have not been fully settled \cite{Mohanty_PRL2012, Dickman_PRE2015, Grassberger_PRE2016}, other problems like characterization of particle transport and dynamic correlations in sandpiles \cite{Pradhan_JSTAT2004, Pradhan_PRE2006, Dickman_EPJB2009, Dickman_JSTAT2014} have not been explored much in the past.
Although understanding spatial and {\it temporal} correlations in various nonequilibrium natural  systems was the original aim of the ``self-organized criticality'' (SOC) hypothesis \cite{Bak_PRL1987},  subsequent research on sandpile models has focused more on the avalanche distributions, and the time-dependent properties of sandpiles have been investigated much less \cite{Jensen_PRB1989, Kertesz-Kiss1990, Manna-Kertesz1991, Bak-Paczuski1995, Dhar-PRE2012}. 
Indeed, even after three decades of intensive studies, there is only a limited knowledge of the large-scale structure of sandpiles in general, and the Manna sandpile in particular. For example, it was realized only recently that the long-ranged correlations in the critical state show hyperuniformity, and its theoretical understanding is still lacking \cite{Levine-PRL2015, Torquato-PRE2016, Grassberger_PRE2016, Dandekar-EPL2020}. Recently we proposed a hydrodynamic theory of conserved stochastic sandpiles \cite{Chatterjee_PRE2018}. Using this theory, here we address the question - ``What is the hydrodynamic time-evolution equation governing density relaxations in the conserved Manna sandpile?''.

Indeed, deriving hydrodynamics of a driven many-body system is difficult in general \cite{Eyink-Lebowitz-Spohn1990, Landim}. Remarkably, hydrodynamics of a special class of sandpile-like models, having a time-reversible dynamics and a steady-state product measure, has been previously derived \cite{Carlson_PRL1990}. 
The Manna sandpile however lacks microscopic time-reversibility and therefore violates detailed balance; consequently, its steady-state measure is not described by the Boltzmann-Gibbs distribution and is a-priori unknown.  Perhaps not surprisingly, a good theoretical understanding of dynamical properties of the Manna sandpile on macroscopic scales is still lacking.

In this paper, we study long-wavelength density relaxations in the one dimensional conserved Manna sandpile, which undergoes an absorbing phase transition below a critical density $\rho_c$. We consider a system on a ring of $L$ lattice sites. We denote local density field, at a lattice point $X$ and time $t$, as $\rho(X, t)$, or, equivalently, the local excess density as $\Delta(X, t) =\rho(X, t) - \rho_c$.   We consider initial density profile $\rho(X, t=0) = g_{in}(X/L)$, where $g_{in}(x)$ with $0 \le x \le 1$ is a piece-wise continuous and smooth function, which describes an initial coarse-grained density profile; unless stated otherwise, we assume periodic boundary condition. We take the limit of large system size $L \rightarrow \infty$  by keeping $g_{in}(x)$ fixed and consider a time-dependent coarse-grained density profile with small wave number $k \rightarrow 0$; we consider cases where correlation length $\xi$ in the system can be either finite or large. In our study, we broadly identify the following regimes of density relaxation.

{\it Regime (1).-- Local density greater than $\rho_c$ everywhere.}  We consider relaxation of local density profile with $\rho(X, 0) > \rho_c$ (or, $\Delta(X, 0) >0$). There are two possibilities: local density is (A) far from criticality ($k \xi \ll 1$) or (B) near criticality  ($k \xi \gsim 1$).

In case (A), density profile is such that $\Delta(X, t) \gsim \rho_c$ and the system is super-critical everywhere. Then, provided an initial profile $\rho(X, t=0) = g_{in}(X/L)$, the time-dependent density profile is of the form $\rho(X, t) = g({X}/{L}, {t}/{L^2})$,
where scaled density field $g(x, \tau)$ satisfies a nonlinear diffusion equation,
\be
\partial_{\tau} g(x, \tau) = \partial_x^2  a(g),
\ee
where $a(g)$ is the coarse-grained density-dependent steady-state activity. This implies that long-wavelength density perturbations having wave numbers $k \rightarrow 0$ relax diffusively, with a finite density-dependent  bulk-diffusion coefficient $D(\rho) = d a/d\rho$ and the relaxation time $\tau_R(k,\rho)$ varies as $\tau_R(k,\rho) \sim k^{-2}/D(\rho)$.

In case (B), the system is invariant under rescaling of position $X\rightarrow \lambda X$, time $t \rightarrow \lambda^z t$, (excess) density $\Delta \rightarrow \lambda^{-\chi} \Delta$ and the bulk diffusivity $D \rightarrow \lambda^{\chi-\alpha} D$, where exponents $z$, $\chi$ and $\alpha$ are the standard critical exponents. Consequently, the transport becomes anomalous as the bulk-diffusion coefficient diverges as $D \sim \xi^{\chi-\alpha}$  and the density profiles with wave numbers $k \rightarrow 0$ relax over a time scale $\tau_R \sim k^{-z}$ with $z = 2+\alpha -\chi <2$. Remarkably, the relaxation time is smaller than that away from criticality. That is, for a fixed $k$, the relaxation time $\tau_R(k,\rho)$ decreases as a function $\rho$ and tends to a non-zero finite value as $\rho \rightarrow \rho_c^{+}$. Note that, for an infinite system, the relaxation time as $k \rightarrow 0$ is still infinite at the critical point.

{\it Regime (2).-- Local density greater than $\rho_c$ in some region and equal to $\rho_c$ elsewhere ($\xi \gg 1$, but $k  \xi \ll 1$).}  In this case, we consider relaxation of a localized density perturbation on infinite critical background.  The system exhibits a self-similar structure in the density range $L^{-1/\nu_{\perp}} \ll \Delta \lsim 1$, where activity $a(\rho) \simeq C \Delta^{\beta}$ has a power-law form with $\beta$ being the order-parameter exponent and $C$ being the model-dependent proportionality constant.
As the system is invariant under rescaling of position $X \rightarrow \lambda X$,  time $t \rightarrow \lambda^{z'}  t$ and (excess) density $\Delta \rightarrow \lambda^{-1} \Delta$, implying the general solution for the time-dependent density
\be
\Delta(X, t) = \frac{1}{(Ct)^{\omega}} {\cal G} \left( \frac{X}{(Ct)^{\omega}} \right),
\ee 
with $\omega=1/z'=1/(1+\beta)$. Here ${\cal G}(y)$ is a function of one variable $y$ and satisfies the differential equation,
\be
- \omega \left[ {\cal G} + y \frac{d \cal G}{d y} \right] = \frac{d^2 {\cal G}^{\beta}}{d y^2}.
\ee
Provided an initial delta-perturbation $\Delta(X, 0)=N_1 \delta(X)$, the above differential equation has a solution
\be
 {\cal G}(y) = \left[ B_0 + B y^2 \right]^{-1/(1-\beta)},
\ee 
where the constants $B_0$ and $B$ can be determined in terms of the exponent $\beta$ and the initially added particle number $N_1$. In this case, the transport is anomalous and characterized by the exponent $z' = 1+ \beta$, which is however different from the dynamic exponent $z$ of regime (1.B);  in general $z'\ne z$ (unless $\nu_{\perp}=1$).

{\it Regime (3).-- Local density greater than $\rho_c$ in some region and less than $\rho_c$ elsewhere.} In this case, initially the density profile is not everywhere greater than critical density. For simplicity, we consider the following initial profile: local density
$$
g_{in}(x) = \rho_c(1+\epsilon_1(x))
$$ 
being greater than critical density in some region and 
$$
g_{in}(x) = \rho_c(1-\epsilon_2(x))
$$ 
being less than critical density elsewhere, with $\epsilon_1, \epsilon_2 > 0$. After some time, when the activity has relaxed locally to the value corresponding to local density, we have the system in a mixed state, made up of active and inactive regions. The active regions then slowly invade the inactive regions and, on large spatio-temporal scales, there is a discontinuity at the boundaries between the active and inactive regions. There are two possible final states: (i) the activity invades all the region and the system eventually becomes homogeneous or (ii) the system becomes frozen (inactive everywhere), where the maximum density is $\rho_c$ and some inactive regions remain uninvaded.

In each of the above regimes, we consider various kinds of initial profiles, which either have a finite number of discontinuities (as in a step-like profile) or vary continuously. In the latter case, density profiles can be smooth (as in a Gaussian profile) or non-smooth (as in a wedge-like profile).
In all cases, our theoretical predictions are in a reasonably good agreement with simulations.

The paper is organized as follows. In Sec. \ref{sec-model}, we define the model of conserved Manna sandpile and, in Sec. \ref{sec-hydrodynamics}, we formulate the hydrodynamic theory for the system. In Sec. \ref{sec-comparison}, we present the detailed predictions of the theory and compare the  theoretical results with simulations; we discuss the following three density regimes -  far-from-critical density regime in Sec. \ref{sec-far-critical}, near-critical density regime in Sec. \ref{sec-near-critical} and the density relaxation on infinite critical background in Sec. \ref{sec-intermediate}. In Sec. \ref{regime3}, we study density relaxation where initial local density is less than critical density in some region and finally we summarize in Sec. \ref{sec-summary}.

\section{Model}
\label{sec-model}

We consider a variant of stochastic Manna sandpiles \cite{Manna_JPhysA1991}  on a one dimensional ring of $L$ sites, where the system evolves in continuous time and total mass is conserved. In the literature, this variant is known as the conserved Manna sandpile (CMS) \cite{Dickman_PRE2001}. Any site $X \in [0, 1, \dots, L-1]$ is assigned an unbounded integer variable, called the number of particles (also called {\it height}) $n_{X} \in [0$, $1$, $2$ $\dots]$. When the particle number $n_X$ at a site $X$ is above a threshold value $n_*=1$, the 

\begin{figure}[H]
\centering
\includegraphics[width=0.5\linewidth]{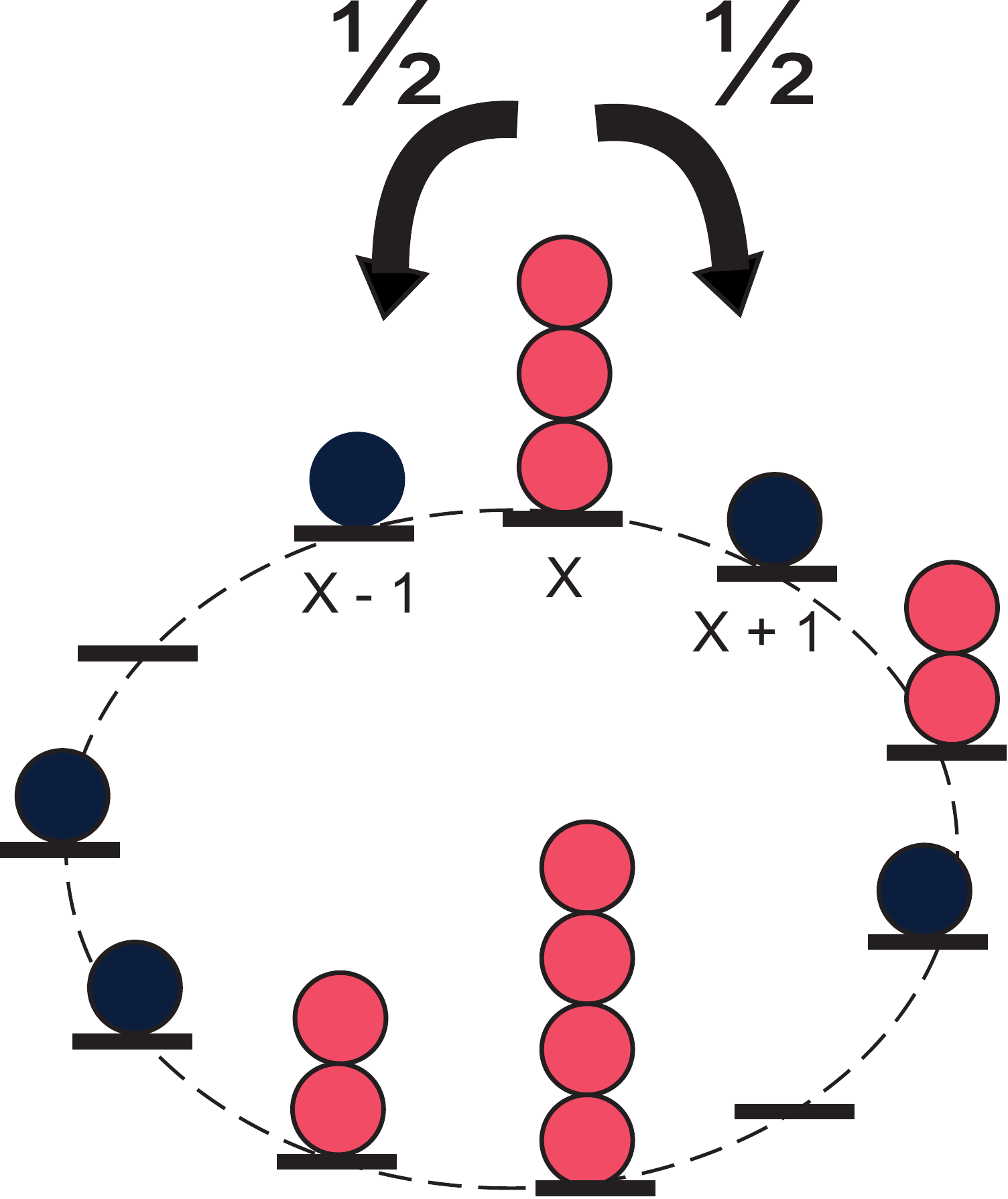}
\caption{{\it Schematic representation of one dimensional conserved manna sandpile.} Particles are represented by full circles; sites are represented by thick black lines and indexed by $X$. The sites having more than one particle are active (red particles with lighter shade); otherwise they (blue particles with darker shade) are inactive. From an active site, two particles are chipped off and each of them are independently transferred to one of the two nearest neighbor sites with equal probability $1/2$.}
\label{schematic-diagram}
\end{figure} 
site is called active. An active site topples by transferring two particles, each of them independently, to the right or the left nearest neighbor sites with equal probability $1/2$; see Fig. \ref{schematic-diagram} for a schematic representation of the update rules. Sites are updated with rate $1$, which sets the time scale in the problem. The total number of particles $N = \sum_{X=0}^{L-1} n_{X}$ remains conserved with density $\rho = N/L$. The activity, which acts as an order parameter for the system, is defined as the average density of active sites 
\be
a(\rho) = \frac{\langle N_{a} \rangle^{st}}{L}
\ee
in the steady state, where $N_a$ is the total number of active sites in the system at a particular time and the average $\langle . \rangle^{st}$ is taken over the steady-state. Note that the steady-state activity $a(\rho)$ is a function of density $\rho$, which is the only tuning parameter in the system. 

Upon tuning the global density $\rho$, the conserved Manna sandpile undergoes an absorbing phase transition below a critical density $\rho_{c}$. Near criticality, the system exhibits critical power-law scaling: when the excess density $\Delta = (\rho - \rho_c) \rightarrow 0^+$ approaches zero from above, the activity $a(\rho)$ vanishes as $a(\rho) \sim \Delta^{\beta}$, with $\beta$ being called the order-parameter exponent, the correlation length and the relaxation time diverge as $\xi \sim \Delta^{-\nu_{\perp}}$ and $\tau_R \sim \Delta^{- \nu_{\perp} z}$, respectively, where $z$ is the dynamical exponent. We shall use the following values of the critical density and the critical exponents: $\rho_c \simeq 0.94885$, $\beta \simeq 0.42$, $\nu_{\perp} \simeq 1.81$ as previously estimated in Ref. \cite{Dickman_PRE2001}.   Notably, as proposed in Ref. \cite{Chatterjee_PRE2018} and demonstrated in this paper, the three exponents $\beta$, $\nu_{\perp}$ and $z$ are not independent, but related through a scaling relation [see eq. (\ref{scaling_relation})].

\section{Hydrodynamics}
\label{sec-hydrodynamics}

Let us consider the conserved Manna sandpile on a ring of $L$ lattice sites. At any lattice point $X$ and time $t$, we can specify the system, on an average level, by defining a local density $\rho(X, t) = \langle n_X(t)\rangle$ - the average of the particle number $n_X(t)$ at site $X$ and time $t$, or, equivalently, by defining a local excess density $\Delta(X, t) =\rho(X, t) - \rho_c$ around the critical density $\rho_c$. 
At time $t=0$, we prepare the system in an initial state with a slowly varying density profile $\rho(X, t=0)$. 
Now, on a lattice of size $L$, the initial density at a site $X$ can be written as a function of  scaled position $x=X/L$,
$$
\rho(X, t=0) \equiv \rho_{in}(X) = g_{in} \left( \frac{X}{L}\right),
$$ 
where the scaled initial density profile $g_{in}(x)$ is a piece-wise continuous and smooth function of the scaled position $x= X/L$. The function $g_{in}(x)$ can have a finite number of discontinuities: it may have a jump in the density value (such as in a step-like density profile) or its derivatives may be discontinuous (such as in a wedge-like density profile) at some points. 
We study relaxation of the initial density profiles, which can be locally in three possible states:
\\
\\
(i) a super-critical state, where local density is greater than the critical density,
\\
\\
(ii) or, a critical  state, where local density is equal to the critical density, 
\\
\\
(iii) or, a sub-critical state, where  local density is less than the critical density.

From the microscopic update rules described in the previous section, we can straightforwardly write down the time-evolution equation for the density field $\rho(X, t)$ at position $X$ and time $t$ as given below \cite{Chatterjee_PRE2018},
\begin{eqnarray}
\frac{\partial \rho(X, t)}{\partial t} &=& [a(X-1,t) - 2 a(X, t) + a(X+1, t)] 
\nonumber \\
&\equiv & \nabla^2 [a(X, t)],
\label{diffusion_discrete}
\end{eqnarray}
where $a(X, t)$ is the average local activity at position $X$ and time $t$ and $\nabla^2$ is the discrete Laplacian. Note that the above time-evolution equation for density field, which is locally conserved, can be written in the form (discrete) of a continuity  equation,
$$
\frac{\partial \rho(X, t)}{\partial t} = J(X, t) - J(X+1,t), 
$$ 
by defining a local current 
\be
J(X,t) = a(X-1, t) - a(X, t) \equiv - \nabla a(X, t)
\label{grad}
\ee 
which is written as a discrete gradient  $\nabla$ of activity $a(X, t)$ and can be identified as the local diffusive current.

Now let us consider the system which is initialized by putting $n_X$ particles at site $X$, where $n_X$ is a Poisson distributed random variable with mean $g_{in}(X/L)$. First we consider the simple case where we randomly distribute particles such that the macroscopic state is homogeneous, i.e., $g_{in}(x) = \rm{const.} = \rho_c (1 + \epsilon)$ with $\epsilon > 0$ finite. Then initially there is a lot of activity, which relaxes to a steady-state value in time $\tau_{act}$. This typical time for relaxation of the activity, in the absence of macroscopic density gradients, is finite. Therefore, we may assume that the activity at all times $t \gg \tau_{act}$ is given by the steady-state value corresponding to the local coarse-grained density, i.e., 
$$a(X, t) = \langle \hat a_X \rangle_{\rho(X, t)}^{st} = a[\rho(X,t)],$$ 
where $\langle . \rangle_{\rho}^{st}$ denotes the steady-state average corresponding to local density $\rho(X,t)$. 
Accordingly, the time-evolution of the density field, from a non-uniform initial profile, is described by a non-linear diffusion equation, 
\be
\frac{\partial \rho(X, t)}{\partial t} = \frac{\partial^2 a(\rho(X, t))}{\partial X^2},
\label{non-lin-diff}
\ee
where $a(\rho)$ is the steady-state activity at density $\rho$. 
In other words, on large spatio-temporal scales, where density field would vary slowly in both space and time,  the coarse-grained local activity is a function of the coarse-grained local density. The above assumption is in the spirit of the assumption of local equilibrium \cite{Eyink-Lebowitz-Spohn1991}, e.g., in the Navier-Stokes hydrodynamic equation, where the equilibrium equation of state connects local pressure to local density and temperature.

Time evolution equation \eqref{non-lin-diff} is invariant under scale transformation $X \rightarrow \lambda X$ and $t \rightarrow \lambda^2 t$.  It implies that the general solution of \eqref{non-lin-diff} has the following scaling form,
\be 
\rho(X, t) = g \left( \frac{X}{L}, \frac{t}{L^2} \right),
\label{diff-scaling}
\ee
and we arrive at the hydrodynamic time-evolution of the scaled density field $g(x,\tau)$,
\begin{equation}
\frac{\partial g(x,\tau)}{\partial \tau} = \frac{\partial^{2} a(g(x,\tau))}{\partial x^{2}} 
\label{diffusion_continuum},
\end{equation}
where rescaled space $x \in [0, 1]$ and time $\tau \in [0, \infty]$ vary continuously in the limit of large $L$ and $a(g)$ is a nonlinear function of the rescaled coarse-grained density $g$. The nonlinear diffusion equation (\ref{diffusion_continuum}) has a unique solution, provided a fixed initial condition $\rho(x, 0) \equiv g_{in}(x)$ and a periodic boundary condition.

The form of the local current as given in eq. \eqref{non-lin-diff} [or, equivalently, eq. \eqref{grad}] helps one to immediately identify the bulk-diffusion coefficient in the system. Writing the time evolution equation (\ref{non-lin-diff}) as a continuity equation for locally conserved density field $\rho(X, t)$,
\be
\frac{\partial \rho(X, t)}{\partial t}=  - \frac{\partial J(\rho(X, t))}{\partial X},
\label{cont-eqn}
\ee 
we obtain local diffusive current given by Fick's law, 
\be
J(\rho) = - \frac{\partial a(\rho)}{\partial X} = - D(\rho) \frac{\partial \rho(X, t)}{\partial X},
\label{J-rho}
\ee 
with the density-dependent bulk-diffusion coefficient
\be 
D(\rho)=\frac{d a(\rho)}{d\rho}.
\label{D-rho}
\ee
Far from criticality, where correlation length $\xi$ is finite, the bulk-diffusion coefficient is also finite and the density perturbations having small wave numbers $k \rightarrow 0$ relax over a time scale $\tau_R \sim k^{-2}/ D(\rho)$ (equivalently,  $\tau_R \sim L^2/D(\rho)$ in a system of size $L$). 
However, when density $\rho \rightarrow \rho_c^+$, the bulk-diffusion coefficient diverges as $D(\rho) \sim (\rho - \rho_c)^{\beta-1}$ because near-critical activity $a(\rho) \sim (\rho-\rho_c)^{\beta}$ has a power-law form with exponent $\beta < 1$ and, consequently, the transport becomes anomalous.

\section{Comparison between hydrodynamic theory and simulations}
\label{sec-comparison}

Indeed it is useful to directly verify the ``local equilibrium'' assumption in simulations of the conserved Manna sandpiles. Accordingly, in this section, we compare the theoretical predictions of eqs. \eqref{non-lin-diff} and \eqref{diffusion_continuum} with microscopic simulations in various regimes of density relaxation.
We consider relaxation of density profiles, which evolve on a large macroscopic scales and are thus typically characterized by small wave numbers $k \ll 1$; also, depending on the density, the correlation length $\xi$ in the system can be finite or large. 
We first consider relaxation of step-like initial density profiles, though other initial conditions, such as Gaussian and wedge-like density profiles, are also studied in a few cases. 


\subsection{Local density greater than $\rho_c$ everywhere}

\subsubsection{Relaxation far from criticality}
\label{sec-far-critical}

Here we study relaxation of the long-wavelength density perturbations where correlation length $\xi$ is finite ($k \xi \ll 1$). That is, in this regime, the system everywhere remains far from criticality with $\Delta (X, t)/ \rho_c \sim {\cal O}(1)$ where excess local density $\Delta(X,t) = \rho(X,t) - \rho_c$.  In simulations, we  generate random initial configurations such that the ensemble average over the configurations corresponds to a given initial density profile. Now, to obtain the density profile at a given final time, we let the system evolve from a particular initial configuration up to that time and perform averaging  over the random initial configurations as well as the stochastic trajectories.

To numerically integrate the hydrodynamic equation \eqref{diffusion_continuum}, we use the standard Euler method, where we discretize space $x$ and time $\tau$ in steps of $\delta x = 10^{-3}$ and $\delta \tau = 10^{-7}$, respectively. We have checked that the integration method conserves the total particle number in the system. Furthermore, as the right hand side of the nonlinear diffusion equation \eqref{diffusion_continuum} is expressed in terms of a nonlinear function $a(\rho)$, we first require to explicitly determine the steady-state (quasi) activity $a(\rho)$ as a function of density $\rho$. The functional form $a(\rho)$ is readily generated through microscopic simulations, where we measure the steady-state activity for various densities, taken in small steps of $\delta \rho = 10^{-2}$. We perform linear interpolation to calculate activity $a(\rho_m)$ at any intermediate density $\rho_{m}$, lying within a density interval $[\rho, \rho + \delta \rho]$.

{\it Relaxation of localized density perturbations on a finite domain.}--
First we consider relaxation of localized density profiles on a uniform background for large position and time in a system with correlation length $\xi$ finite ($k \xi \ll 1$).
We compare the time-evolved density profiles obtained by integrating the hydrodynamic  equation \eqref{diffusion_continuum} and that obtained from Monte Carlo simulations for step-like initial localized density  profile, 
\begin{eqnarray}
\rho_{in}(X) =\left\lbrace
\begin{array}{ll} 
\rho_1 + \rho_0 & \mbox{for} ~ \frac{L}{2} - \frac{\delta}{2} <  X < \frac{L}{2} + \frac{\delta}{2}, \cr
\rho_0 & \mbox{otherwise,}           
\end{array}
\right.
\label{step-loc}
\end{eqnarray}
where the step height is $\rho_1 = 4.0$. We take system size $L=1000$ and the width of the profile is chosen to be $\delta =40$, where $\delta/L \ll 1$ being small, implying that the initial density perturbation is well localized on the macroscopic scale. The initial density perturbation is generated by adding $N_1=160$ particles over a uniform background having density  $\rho_{0}=1.0$. Note that, throughout the paper, we denote the global density as $\bar \rho$, which, for large $L$, can be written as the spatially averaged scaled local density, 
$$\bar{\rho} = \frac{1}{L} \sum_{X=1}^{L} \rho(X, t) \equiv \int_{0}^{1} g(x,\tau) dx.$$
 In this particular case, we take the global density $\bar \rho = 1.16$.

\begin{figure}[H]
\centering
\includegraphics[width=1.0\linewidth]{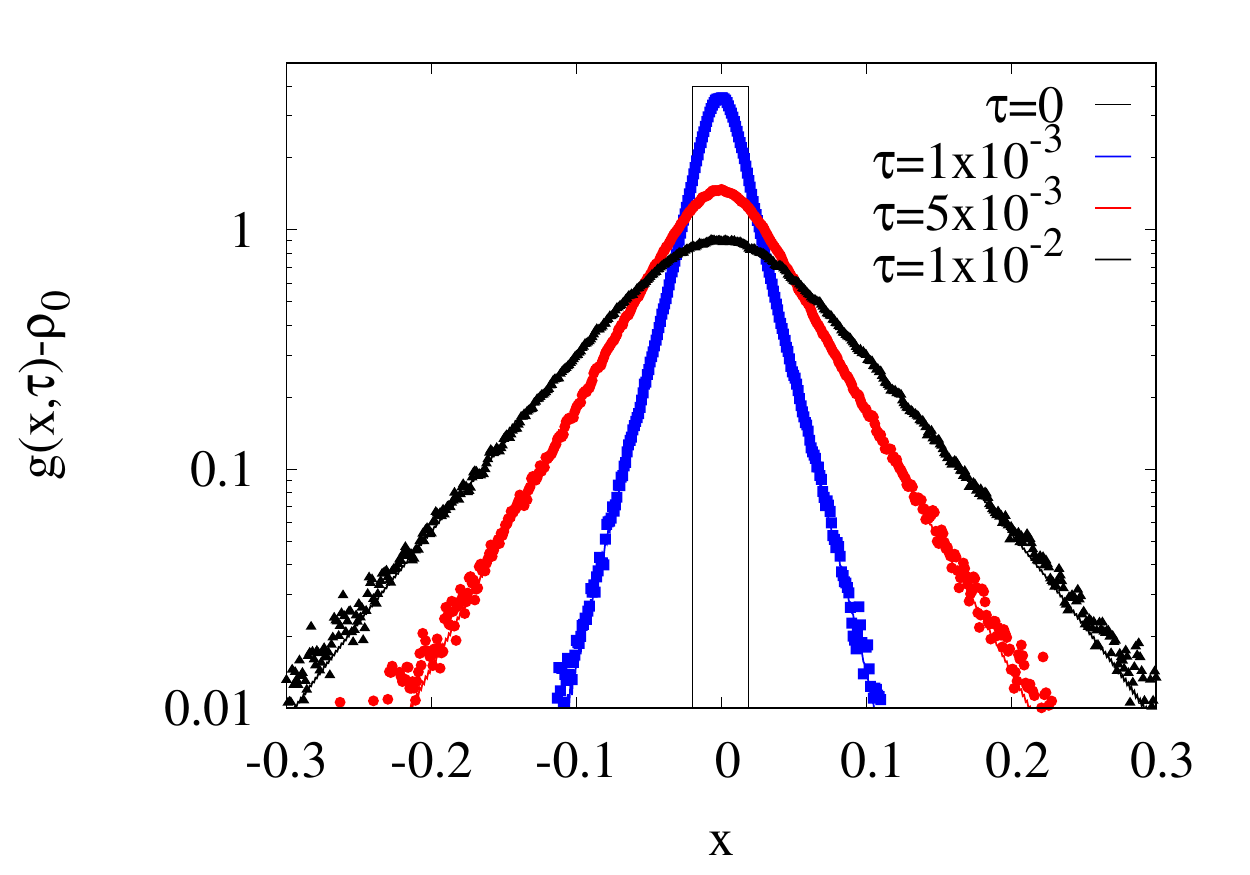}
\caption{{\it Verification of eq. \eqref{diffusion_continuum} for localized step-like initial density perturbations on a uniform background.}  Scaled density (shifted) field $g(x,\tau) - \rho_0$, measured around a uniform background density $\rho_0=1$, is plotted as a function of shifted scaled position variable $x$ where we set the origin $x=0$ such that the peak is centered at the origin. We plot the density profiles at various hydrodynamic times $\tau = 10^{-3}$ (blue squares), $5 \times 10^{-3}$(red circles) and $10^{-2}$ (black triangles), where lattice position $X$ and time $t$ are scaled as $x=X/L$ and $\tau=t/L^2$. We consider  a step-like initial profile Eq. \eqref{step-loc}. System size $L=1000$ and global density $\bar \rho=1.16$; lines - theory [numerically integrated Eq. \eqref{diffusion_continuum}], points - Monte Carlo simulations.}
\label{fig-loc}
\end{figure}

In Fig. \ref{fig-loc}, we plot scaled shifted density field $g(x,\tau) - \rho_0$, measured around background density $\rho_0$, as a function of the shifted rescaled position $x$ at various hydrodynamic times $\tau = 10^{-3}$ (blue squares), $5 \times 10^{-3}$ (red circles) and $10^{-2}$ (black triangles) for the step-like initial profile eq. \eqref{step-loc}; lattice position $X$ and time $t$ in simulations are related to  hydrodynamic (shifted) position and hydrodynamic time $x=X/L-1/2$ and $\tau=t/L^2$, respectively; in simulations, we perform averages over $2 \times 10^4$ random initial configurations and trajectories to obtain density profiles at the final times.  We numerically integrate the nonlinear diffusion equation \eqref{diffusion_continuum}, using the Euler method, up to hydrodynamic times  $\tau = 10^{-3}$ (blue lines), $5 \times 10^{-3}$(red lines) and $10^{-2}$ (black lines), from the initial condition \eqref{step-loc}; in Fig. \ref{fig-loc}, we also plot the numerically integrated shifted scaled density profiles $g(x, \tau) -\rho_0$ as a function of scaled position $x$ for various hydrodynamic times $\tau$. One can see that the density profiles obtained from numerically integrated eq. \eqref{diffusion_continuum} (lines) is in quite good agreement with that obtained from simulations (points), almost over a couple of decades of the density values.
We have also studied other initial profiles, such as wedge-like and Gaussian profiles, and find a nice agreement between hydrodynamic theory and simulations (not presented here). As the width of the initial profiles in all cases are  small compared to the system size, the initial density profile on the hydrodynamic scale can be approximated as the Dirac-delta function; consequently the time-evolved profiles at large times are independent of the exact shape of the initial profiles and depend only on the strength of the delta function (i.e., the initially added number of particles).

\begin{figure}
\centering
\includegraphics[width=1.0\linewidth]{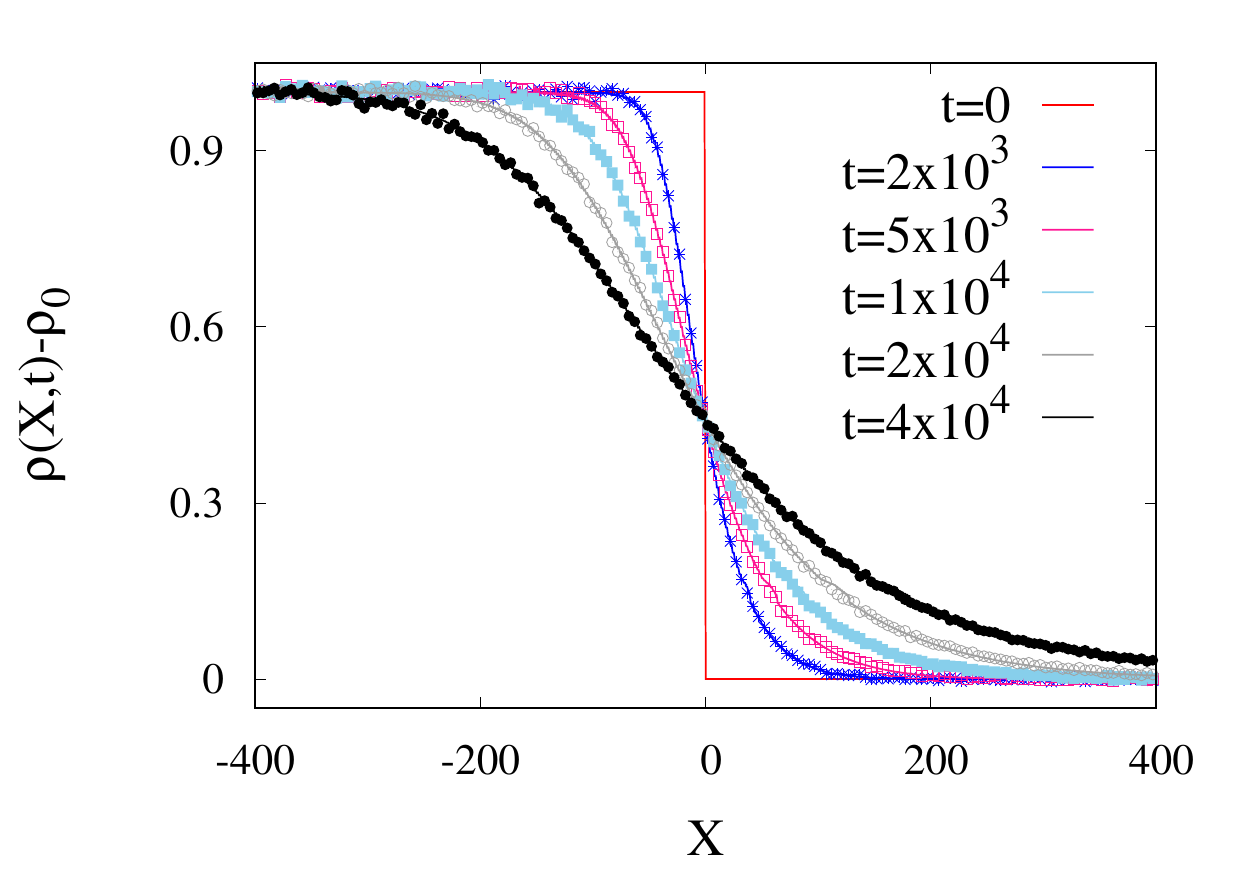}
\caption{ {\it Base density of initial step profile greater than the critical density ($\rho_0 > \rho_c$).}   Excess densities, evolved from step initial condition Eq. \eqref{step-init} in simulations, are plotted as a function of position $X$ at various times $t=2 \times 10^3$ (blue asterisks), $5 \times 10^3$ (pink open squares), $10^4$ (sky-blue filled squares), $2 \times 10^4$ (grey open circles) and $4 \times 10^4$ (black filled circles). We choose $\rho_0=1$ and $\rho_1=2$ in the initial profile eq. \eqref{step-init}. }
\label{abv-rhoc}
\end{figure}

{\it Relaxation of step profile on infinite super-critical background.}--
Next we consider relaxation of a step-like density profile spreading on infinite super-critical background. We create an initial density perturbation, on the left half of the origin $X=0$, in the form of steps over a uniform background having density $\rho_0=1.0 > \rho_c$ and study how the density perturbation propagates into the domain on the right side of the origin. The initial step-like density profile is given by
\begin{eqnarray}
\rho_{in}(X) =\left\lbrace
\begin{array}{ll} 
\rho_{1} + \rho_{0} & \mbox{for} ~ -\infty < X \leq  0, \cr
\rho_{0} & \mbox{otherwise,}           
\end{array}
\right.
\label{step-init}
\end{eqnarray}
where $\rho_{1}$ and $\rho_{0}$ are the height and the base density of the step profile, respectively.

As the local density in this case remains far from criticality ($\xi$ finite), the diffusion coefficient remains finite throughout in the space and time domain considered here and the transport is therefore diffusive in nature.  In Fig. \ref{abv-rhoc}, we plot the shifted density profile $\rho(X,t)-\rho_0$ as a function of position $X$ at various times  $t=2 \times 10^3$ (blue asterisks), $5 \times 10^3$ (pink open squares), $10^4$ (sky-blue filled squares), $2 \times 10^4$ (grey open circles) and $4 \times 10^4$ (black filled circles).  We then compare the density profiles obtained from simulations with that obtained by numerically integrating the nonlinear diffusion equation \eqref{non-lin-diff}. We find excellent agreement between hydrodynamic theory (lines) and simulations (points).

{\it Verification of diffusive scaling of eqs. \eqref{diff-scaling} and \eqref{diffusion_continuum}.}--
In this section, we directly verify the diffusive scaling limit, which has been used to obtain the time evolution equation \eqref{diffusion_continuum}. In this scaling regime, we plot the scaled time-dependent density profiles $\rho(X=x L, t = \tau L^2) \equiv g(x, \tau)$ as a function of the scaled position $x = X/L$ for different system sizes $L$ and different times $t$ by keeping the hydrodynamic time $\tau = t/L^{2}$ fixed. Then, according to the diffusive scaling as in \eqref{diff-scaling}, different curves should collapse onto each other. Moreover, the collapsed profiles at the fixed hydrodynamic time $\tau$ should be described by the nonlinear diffusion equation \eqref{diffusion_continuum}, integrated up to time $\tau$ from a given initial density profile.

\begin{figure}[H]
\centering
\includegraphics[width=0.98\linewidth]{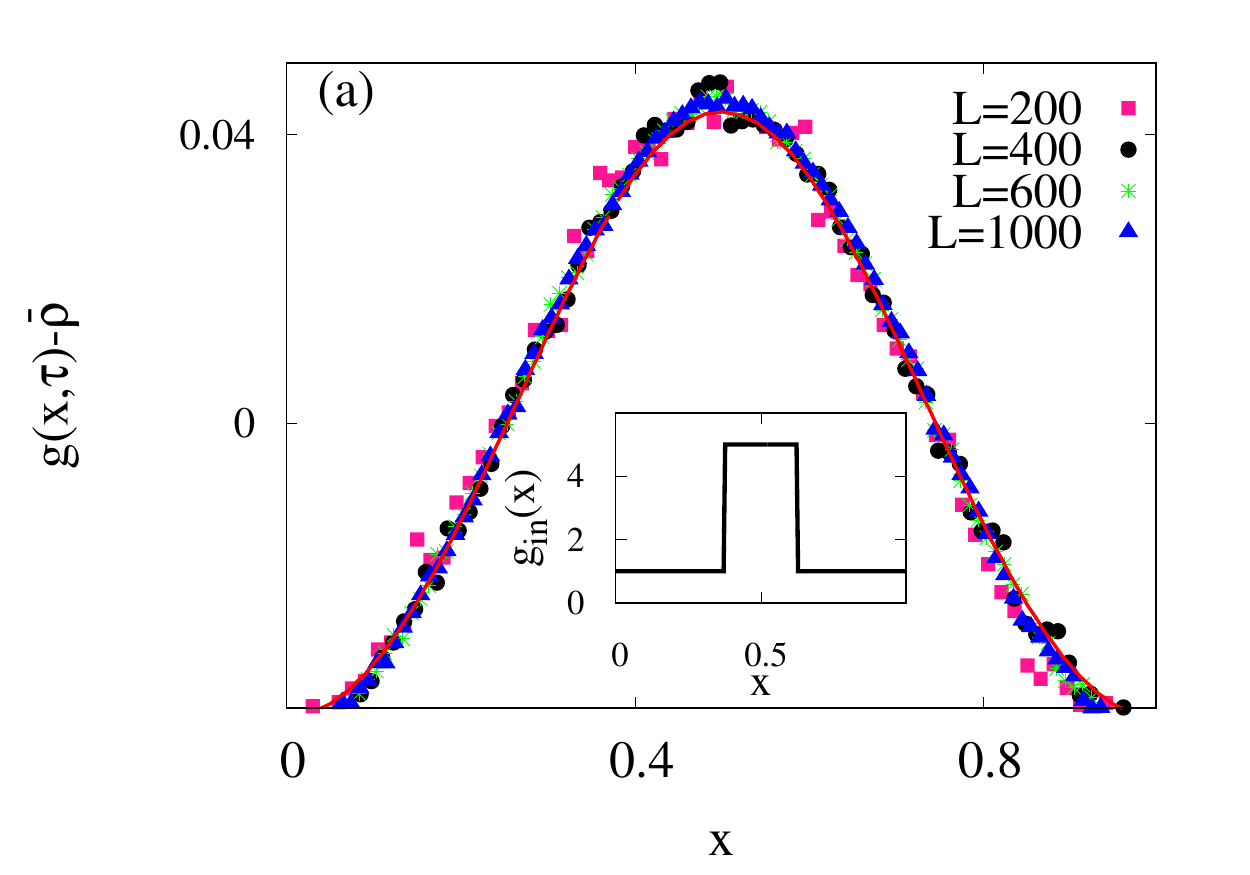}
\includegraphics[width=0.98\linewidth]{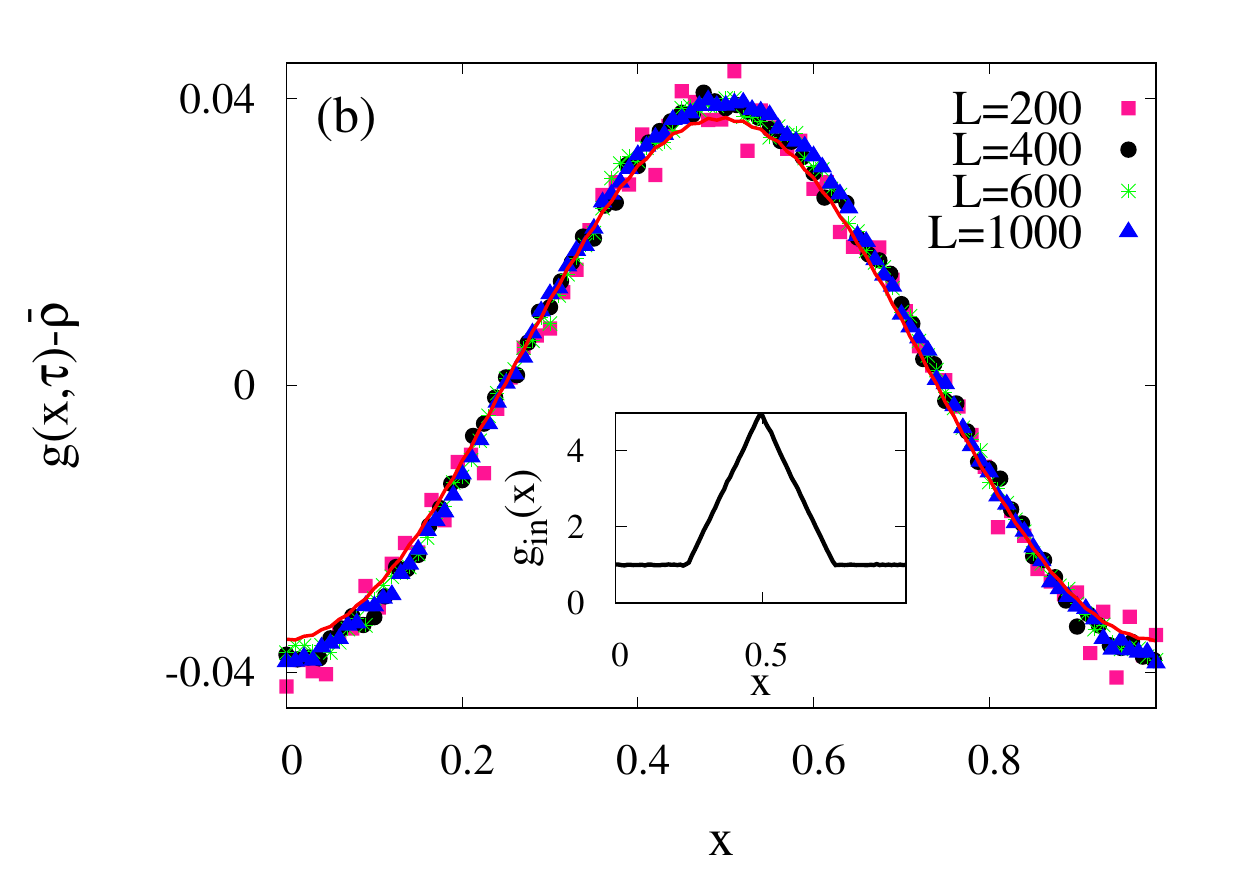}
\caption{ {\it Verification of diffusive scaling in eq. \eqref{diffusion_continuum}.} Scaled density profile $g(x,\tau)-\bar \rho$, measured around the spatially averaged density $\bar \rho$, is plotted as a function of rescaled position $x=X/L$ at different  times:   $t=2 \times 10^4$ for $L=200$ (pink squares), $t=8 \times 10^4$ for $L=400$ (black circles), $t=1.8 \times 10^5$ for $L=600$ (green asterisks) and $t= 5 \times 10^5$ for $L=1000$ (blue triangles), by keeping the hydrodynamic time $\tau=t/L^2=0.5$ fixed. Simulation points at different times collapse onto each other reasonably well. 
We take two sets of initial density perturbations $g_{in}(x)$, with $0 \le x \le 1$ and periodic boundary condition, over a uniform background having density $\rho_{0}=1.0$: step-like perturbation of height $4.0$ [Eq. \eqref{step-diff}] in panel (a) and wedge-like perturbation of width $1/2$ [Eq. \eqref{wedge-diff}] in panel (b);  spatially averaged (global) density $\bar{\rho}=2.0$ is same for all plots. 
Insets: The initial scaled density profiles $g_{in}(x)$ are plotted as a function of scaled position $x$. Lines - theory [numerical integration of eq. \eqref{diffusion_continuum}], points - Monte Carlo simulations.}
\label{fig-diff}
\end{figure}

We now check the above assertions in simulations for two different initial conditions - step-like and wedge-like profiles and for various system sizes $L=200$, $400$, $600$ and $1000$. The scaled step-like initial profile is chosen as
\begin{eqnarray}
\rho \left(\frac{X}{L}=x, t=0 \right) \equiv g_{in}(x) =\left\lbrace
\begin{array}{ll} 
\rho_{1} + \rho_{0} & \mbox{for} ~ 0<x<x_{1}, \cr
\rho_{0} & \mbox{otherwise.}           
\end{array}
\right.
\label{step-diff}
\end{eqnarray}
The above profile is generated by distributing $N_1=L(\bar \rho - \rho_0)$ number of particles in a box of width $x_{1}=1/4$ and height $\rho_{1} = 4.0$ over a uniform background having density  $\rho_{0} = 1.0$ so that the global density $\bar{\rho} = 2.0.$ A wedge-like initial profile is chosen as
\begin{eqnarray}
g_{in}(x) =\left\lbrace
\begin{array}{ll} 
\rho_0 + 2 \rho_1(x - x_1)/\delta & \mbox{for} ~ x_1 < x < x_2, \cr
\rho_0 + 2 \rho_1(x_3 - x)/\delta & \mbox{for} ~ x_2 < x < x_3, \cr
\rho_0 & \mbox{otherwise.}           
\end{array}
\right.
\label{wedge-diff}
\end{eqnarray}
The wedge is centered around $x=x_2=1/2$, has a width $\delta=1/2$, ranging from $x=x_1=(1- \delta)/2$ to $x=x_3=(1+\delta)/2$, and has a height $\rho_1=4.0$. As in the previous case, $N_1=L(\bar \rho - \rho_0)$ number of particles are distributed over the uniform background having density $\rho_{0}=1.0$, by keeping global density $\bar{\rho} = 2.0$.
In simulations, we take the final hydrodynamic time $\tau=0.5$ and each system is allowed to evolve up to time $t= \tau L^2$, depending on the respective system size $L$. 
In Fig. \ref{fig-diff}, we plot the scaled density profile $g(x,\tau)-\bar \rho$, measured around the spatially averaged (global) density $\bar \rho = \int_0^1 g(x, \tau) dx$, as a function of rescaled position $x=X/L$ at different times:  $t=2 \times 10^4$ for $L=200$ (pink squares), $t=8 \times 10^4$ for $L=400$ (black circles), $t=1.8 \times 10^5$ for $L=600$ (green asterisks) and $t=5 \times 10^5$ for $L=1000$ (blue triangles); we perform simulations for the step-like initial profile [eq. \eqref{step-diff}] in panel (a) and for the wedge-like initial profile [eq. \eqref{wedge-diff}] in panel (b) of the figure. We perform averages over $2 \times 10^5$ random initial configurations and trajectories. We observe reasonably good scaling collapse of all density profiles, obtained at four different times and system sizes. Next, we numerically integrate the time evolution  equation \eqref{diffusion_continuum}, from the above two initial conditions as in eqs. \eqref{step-diff} and\eqref{wedge-diff}, up to hydrodynamic time $\tau=0.5$, using the Euler method and plot in Fig. \ref{fig-diff} the scaled density $g(x, \tau=0.5)$ as a function of scaled position $x$. Indeed, we find the hydrodynamic theory (red lines) in reasonably good agreement with the collapsed density profiles obtained from simulations (points).

\subsubsection{Near-critical density relaxation}
\label{sec-near-critical}

Here we study  relaxation of  long-wavelength density perturbation having wave numbers $k \rightarrow 0$ in a system where correlation length $\xi \gg 1$ is also large, keeping $k \xi$ finite. This regime essentially corresponds to the local density $\rho(X, t) - \rho_c \equiv \Delta(X, t) \sim L^{-1/\nu_{\perp}}$ very small such that the correlation length becomes of order system size ($\xi  \sim L$). In this case, the hydrodynamic equation \eqref{diffusion_continuum}, where we have used diffusive time-scaling, is not expected to hold and therefore it cannot describe density relaxation in the system.  The physical origin of the break down of the diffusive scaling is perhaps not difficult to understand. The activity, which behaves near criticality as a power law $a(\rho) \sim \Delta^{\beta}$ with $\beta < 1$, has a singularity at the critical point $\Delta=0$ and, consequently, its derivative with respect to the density diverges, leading to a diverging bulk-diffusion coefficient $D(\Delta) \sim \Delta^{\beta-1} \sim \xi^{(1-\beta)/\nu_{\perp}}$ [see eq. \eqref{D-rho}]. That is, the transport in the system near criticality becomes ``super-diffusive'' or anomalous.

Indeed one can make the argument more precise in terms of a finite-size scaling analysis. In the conserved Manna sandpiles near criticality, where $\Delta \sim L^{-1/\nu_{\perp}} \ll 1$, the activity in the system is known to have the following finite-size scaling form \cite{Dickman_PRE2001},
\begin{equation}
a(\Delta,L) = L^{-\beta/\nu_{\perp}}\mathcal{A}(L^{1/\nu_{\bot}}\Delta),
\label{activity_exp}
\end{equation}
where $\beta$ is the order-parameter exponent, the exponent $\nu_{\bot}$ characterizes the diverging correlation length, and $\mathcal{A}(L^{1/\nu_{\bot}}\Delta)$ is the scaling function; note that, for large $y \gg 1$,  ${\cal A}(y) \sim y^{\beta}$ and we recover the power law scaling of activity $a(\Delta) \sim \Delta^{\beta}$. 
Accordingly, the near-critical bulk-diffusion coefficient has the following scaling form 
\be
D(\Delta, L) = \frac{d a}{d \Delta} = L^{(1-\beta)/\nu_{\perp}} \mathcal{A}^{\prime}(L^{1/\nu_{\perp}}\Delta).
\label{D-critical}
\ee
Now, by substituting eq. \eqref{D-critical} into the continuity equation \eqref{cont-eqn} through eqs. \eqref{J-rho} and \eqref{D-rho} and then changing variable from density $\rho(X, t)$ to excess density $\Delta(X, t) = \rho(X, t)-\rho_c$, we have
\begin{equation}
\frac{\partial \Delta(X,t)}{\partial t} = \frac{\partial}{\partial X} \left[ L^{(1-\beta)/\nu_{\perp}} \mathcal{A}^{\prime}(L^{1/\nu_{\perp}}\Delta) \frac{\partial \Delta}{\partial X}\right].
\label{super-diffusion-no-t-scaling}
\end{equation}
Indeed, the time-evolution equation \eqref{super-diffusion-no-t-scaling} can be recast in a compact form,
\begin{equation}
\frac{\partial G(x,\tau)}{\partial \tau} = \frac{\partial^{2}\mathcal{A}(G)}{\partial x^2},
\label{hydro-super-diff}
\end{equation}
by performing scale transformation of excess density $\Delta$, position $X$ and time $t$ to a rescaled excess density $G(x,\tau)$, rescaled position $x$ and rescaled time $\tau$, respectively, as
\bea
G(x,\tau) &=& L^{1/\nu_{\perp}} \Delta(X, t),
\label{sd-limit1} \\
x &=& \frac{X}{L},
\label{sd-limit2} \\
\tau &=& \frac{t}{L^z},
\label{sd-limit3}
\eea 
where the dynamical exponent $z$ can be expressed in terms of the two static exponents $\beta$ and $\nu_{\bot}$, through a scaling relation
\be 
z=2 - \frac{1-\beta}{\nu_{\perp}}.
\label{scaling_relation}
\ee 
The above relation was first proposed in Ref. \cite{Chatterjee_PRE2018}. However, the finite-size scaling arguments given above, and the subsequent derivation of the time-evolution equation \eqref{hydro-super-diff} for the near-critical density field, are new to the best of our knowledge and can be directly tested in simulations.\\~\\
We verify the hydrodynamic time-evolution of the rescaled excess density field $G(x,\tau)$ by comparing the time-evolved density profile obtained from numerically integrating  eq.\eqref{hydro-super-diff} and that obtained from simulations. For the purpose of the numerical integration of eq. \eqref{hydro-super-diff}, one needs to know ${\cal A}(G)$ as a function of  $G$. The scaling function ${\cal A}(G)$ was previously calculated numerically in Ref. \cite{Dickman_PRE2001} and can be quite well described by the functional form ${\cal A}(G) = c_0 (1 + c_1 G)^{\beta}$, where $c_0 \simeq 0.18$ and $c_1 \simeq 7.0$. This particular form can be understood from the fact that the power-law dependence of the near-critical activity $a \propto \Delta^{\beta}$ is cut off at a very small excess density $\Delta \sim L^{-1/\nu_{\perp}}$. Consequently, as the excess density (or, density) approaches zero (critical density), the scaling function ${\cal A}$ does not actually vanish, and takes a finite value $c_0$. 
In the numerical integration and simulations, we consider a step-like initial profile $G_{in}(x) \equiv G(x,\tau =0)$ for the rescaled excess density,
\begin{eqnarray}
G_{in}(x) =\left\lbrace
\begin{array}{ll} 
\rho_{1}  & \mbox{for} ~  x_1 \leq x \leq x_2, \cr
0 & \mbox{otherwise.}           
\end{array}
\right.
\label{step-super-diff}
\end{eqnarray}
Here we set $x_1 = 1/2-\delta/2$ and $x_2=1/2 + \delta/2$, with $\rho_1 \simeq 6.69$ and $\delta = 1/4$ being the height and the width of the initial step profile, respectively; the profile is created over a uniform critical background having density $\rho_{c} \simeq 0.94885$.  
In simulations, to generate the above initial density profile, we uniformly distribute an appropriate number of particles in a domain of size $L/4$, over uniform critical background configurations. We generate critical background configurations by using a standard algorithm for generating one dimensional quasi-periodic strings of $0$'s and $1$'s of size $L$, by ensuring that the system has a fixed background density $\rho_c$ \cite{quasi-periodic}. 

First we verify the ``super-diffusive'' scaling as in eqs. \eqref{sd-limit1}, \eqref{sd-limit2} and \eqref{sd-limit3}, which have been used to obtain the hydrodynamic time-evolution equation \eqref{hydro-super-diff}. In this way, we can test the scaling relation as given in eq. \eqref{scaling_relation}. We take four systems of sizes  $L_1=1500$, $L_2=2000$, $L_3=2500$ and $L_4=5000$, which are allowed to evolve from the step-like initial density profile eq. \eqref{step-super-diff} up to times $t_1=\tau L_1^z$, $t_2 = \tau L_2^z$, $t_3 = \tau L_3^z$ and $t_4=\tau L_4^z$, respectively, with $\tau$ fixed;  here the value of the dynamic exponent $z \simeq 1.68$ is  calculated using the scaling relation eq. \eqref{scaling_relation}, where we use the previously estimated values of the exponents $\beta \simeq 0.42$ and $\nu_{\perp} \simeq 1.81$ for the conserved Manna sandpile \cite{Dickman_PRE2001}. According to the time scaling in eq. \eqref{sd-limit3}, the above three time-evolved density profiles should collapse onto each other. 

\begin{figure}[H]
\includegraphics[width=1.0\linewidth]{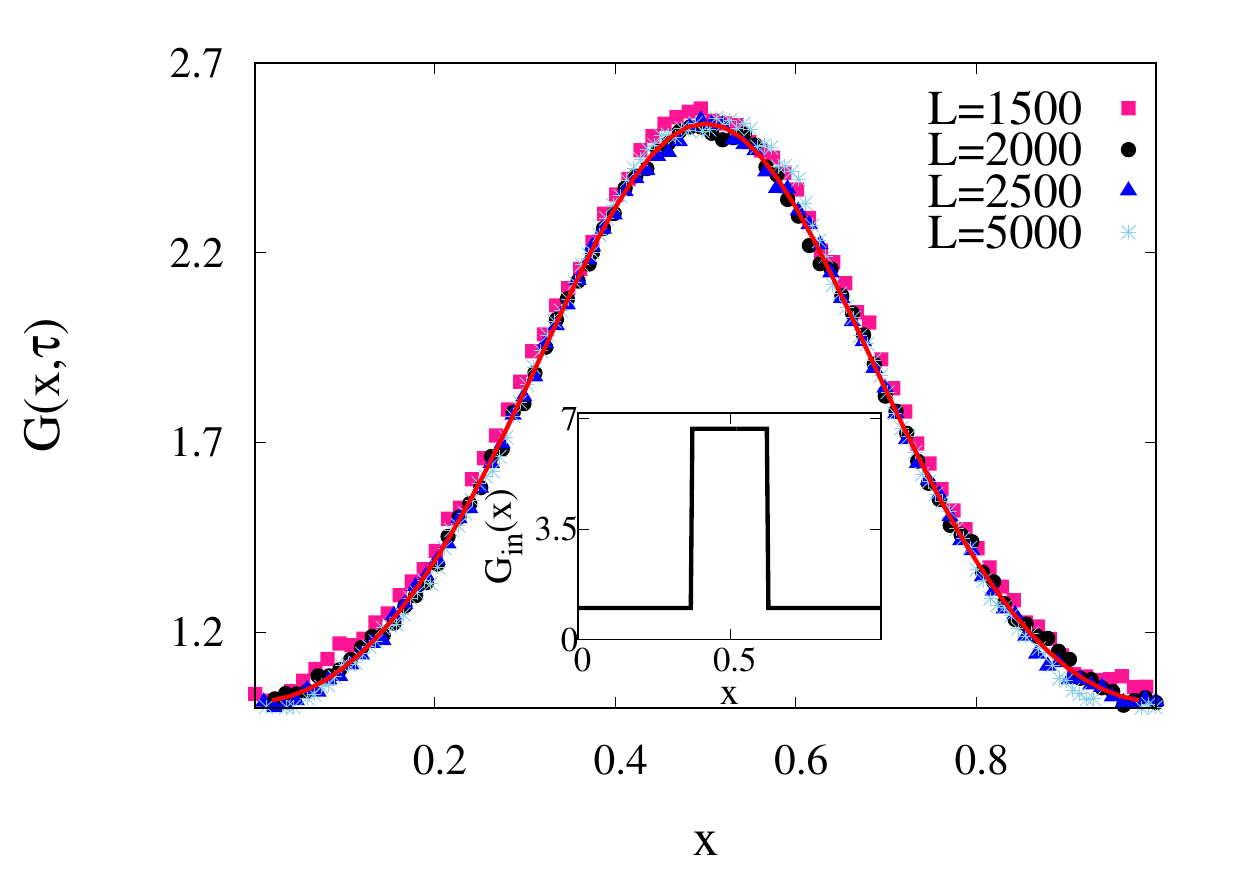}
\caption{ {\it Verification of  anomalous scaling in eq. \eqref{hydro-super-diff}.} Scaled local excess density $G(x, \tau) \equiv L^{1/\nu_{\perp}} \Delta(X=x L, t= \tau  L^z)$ is plotted as a function of scaled position $x$  at different times and system sizes:  $t = 59286$ for $L=1500$ (pink squares), $t = 96128$ for $L=2000$ (black circles), $t = 139849$ for $L=2500$ (blue triangles) and $t=448116$ for $L=5000$ (sky-blue asterisks), where we have kept the hydrodynamic time $\tau=t/L^z \simeq 0.27$ fixed; here we have calculated $z \simeq 1.68$ by using eq. \eqref{scaling_relation}, where we have used $\beta \simeq 0.42$ and $\nu_{\perp} \simeq 1.81$ \cite{Dickman_PRE2001}. Points obtained from simulations at different times and system sizes collapse onto each other reasonably well. The red line corresponds to the numerical integration of eq. \eqref{hydro-super-diff}. 
Inset: Initial scaled excess density profile $G_{in}(x) \equiv L^{1/\nu_{\perp}} \Delta(X=x L, t=0)$ is plotted as a function of scaled position $x=X/L$.
}
\label{fig6:finite_size_scaling}
\end{figure}

In Fig. \ref{fig6:finite_size_scaling}, we plot the scaled excess density $G(x,\tau) \equiv L^{1/\nu_{\perp}} \Delta(X, t)$ as a function of the scaled position $x=X/L$ for the four systems at the following times: $t_1 = 59286$ for $L_1=1500$ (pink squares), $t_2 = 96128$ for $L_2=2000$ (black circles), $t_3 = 139849$ for $L_3=2500$ (blue triangles) and $t_4=448116$ for $L=5000$ (sky-blue asterisks),  where we take $\tau=t/L^z \simeq 0.27$; we perform averages over $2 \times 10^5$ random initial configurations and trajectories. We observe a quite good scaling collapse of the final scaled excess density profiles. Also, according to the theory, the collapsed density profiles should be described by the time-evolution equation \eqref{hydro-super-diff}, which is integrated up to hydrodynamic time $\tau \simeq 0.27$ using the initial condition as in eq. \eqref{step-super-diff}. In Fig. \ref{fig6:finite_size_scaling}, the numerically integrated scaled excess density $G(x,\tau)$ at $\tau \simeq 0.27$ is plotted as a function of position $x$ (red line); in inset, scaled initial excess density profile $G(x, \tau=0) \equiv G_{in}(x)$ is plotted as a function of scaled position $x$. We find that the theory (red line) is in a quite agreement with simulations (points).

\subsection{Local density greater than $\rho_c$ in some region and equal to $\rho_c$ elsewhere}
\label{sec-intermediate}

Here we consider relaxation of density profile on infinite critical background having density $\rho_c$. We initially create a density perturbation in some region (finite or infinite), where the local density at $t=0$ is greater than the critical density; at later times $t>0$, the density profiles start spreading over the critical background. We study below the time evolution of the profiles on large spatio-temporal scales.

\subsubsection{Relaxation of localized density perturbation on infinite critical background}

First we study the time evolution of an initially localized density profile on infinite critical background.  Indeed, in this case, we can analytically calculate the asymptotic scaling form of the time-dependent density profile.  We study relaxation of the localized density perturbation on large space and time scales and the system is not far from criticality (i.e., in the long-wavelength regime where $\xi \gg 1$, but $k \xi \ll 1$). Here the time-evolved excess density field $\Delta(X, t) = (\rho(X, t) - \rho_{c})$ is still quite small, but much larger than ${\cal O}(L^{-1/\nu_{\perp}})$. In other words, our analysis is valid for the relaxation of density profiles, for which the local excess density $\Delta(X,t)$ remains in the range of $L^{-1/\nu_{\perp}} \ll \Delta \lsim 1$. In this regime, the activity is known to have a power-law dependence on excess density and is given as, 
\be 
a(\Delta) \simeq C \Delta^{\beta},
\label{power-law-a}
\ee
where $C$ is a model-dependent proportionality constant. 
Now, by substituting the above power-law form of the activity in the diffusion equation \eqref{non-lin-diff} and changing variable to $\Delta(X, t) = \rho(X,t) +\rho_c$, we obtain the time-evolution equation for the excess density field $\Delta(X,t)$,
\begin{equation}
\frac{\partial \Delta(X, t)}{\partial t} = C\frac{\partial^2 [\Delta(X,t)]^{\beta}}{\partial X^2}.
\label{diffusion_equation_putting_activity}
\end{equation}
We note that the above equation is invariant under a scale transformation of position, time and density field,
\bea
X \rightarrow \lambda X, 
\nonumber \\
t \rightarrow \lambda^{z'}  t,
\nonumber \\
\Delta \rightarrow \lambda^{-\chi'} \Delta,
\label{scale-trans}
\eea
 provided that the following relation is satisfied,
\be
\chi' + z' = \chi' \beta + 2.
\label{scaling1}
\ee 
Here we are interested in studying how a localized initial density perturbation having a finite width would relax on infinite critical background on large spatio-temporal scales. To this end, we specifically consider the Dirac-delta initial condition 
\be
\Delta(X, t=0) = N_1 \delta(X),
\label{delta-init-cond}
\ee
where $N_1$ is the strength of the delta function, i.e., the number of particles added at the origin as an initial perturbation over the critical background. Interestingly, in this case, the time-dependent density profile $\Delta(X, t)$, which is governed by the nonlinear diffusion equation \eqref{diffusion_equation_putting_activity}, can be exactly solved, provided that the solution satisfies the boundary condition $\Delta(x = \pm \infty, t) = 0$. 
Due to the scale-invariant structure of eq. \eqref{diffusion_equation_putting_activity}, one would naturally expect a scale invariant solution for the excess density $\Delta(X, t)$. We proceed with the following scaling ansatz, 
\begin{equation}
\Delta(X, t) = \frac{1}{(C t)^{\omega}} {\cal G} \left[\frac{X}{(C t)^{\omega}}\right],
\label{scaling-fn}
\end{equation}
where ${\cal G}(y)$ is the scaling function of a single scaling variable $y=X/(Ct)^{\omega}$. Now, under the scale transformation of eq. \eqref{scale-trans}, the density field and the scaling variable in eq. \eqref{scaling-fn} transform  as $\Delta \rightarrow \lambda^{-z' \omega} \Delta$ and $y \rightarrow \lambda^{1-z' \omega} y$, respectively. Now, by identifying $z' \omega = \chi' $, putting $1- z' \omega = 0$ (which leaves the scaling variable $y$ invariant) and substituting $z' = 1/\omega$ and $\chi' = 1$ in eq. \eqref{scaling1}, we exactly determine the growth exponent
\be
\omega=\frac{1}{1+\beta}.
\label{scaling2}
\ee
Moreover, the differential equation satisfied by the scaling function ${\cal G}(y)$ can be straightforwardly  written as
\be
-\omega \left[ {\cal G} + y\frac{d {\cal G}}{dy} \right] = \frac{d^2 {\cal G}^{\beta}}{dy^2}, 
\label{scaling-soln}
\ee
which, for the delta function initial perturbation [Eq. (\ref{delta-init-cond})] on infinite background, has the following solution
\begin{equation}
{\cal G}(y) = \frac{1}{\left[ g_0^{\beta - 1} + \frac{\omega(1 - \beta)}{2\beta} y^2 \right]^{1/(1 - \beta)}}.
\label{gy}
\end{equation}
Here we have used the boundary conditions
\begin{eqnarray}
{\cal G}(y = 0) = g_{0}; ~~
\left[ \frac{d {\cal G}}{dy}\right]_{y=0} = 0,
\end{eqnarray}
where the constant $g_0$ is fixed by the normalization condition $\int^{\infty}_{-\infty} g(y) dy = N_1$, the number of particles added in the system at $t=0$. 

\begin{figure}[H]
\includegraphics[width=1.0\linewidth]{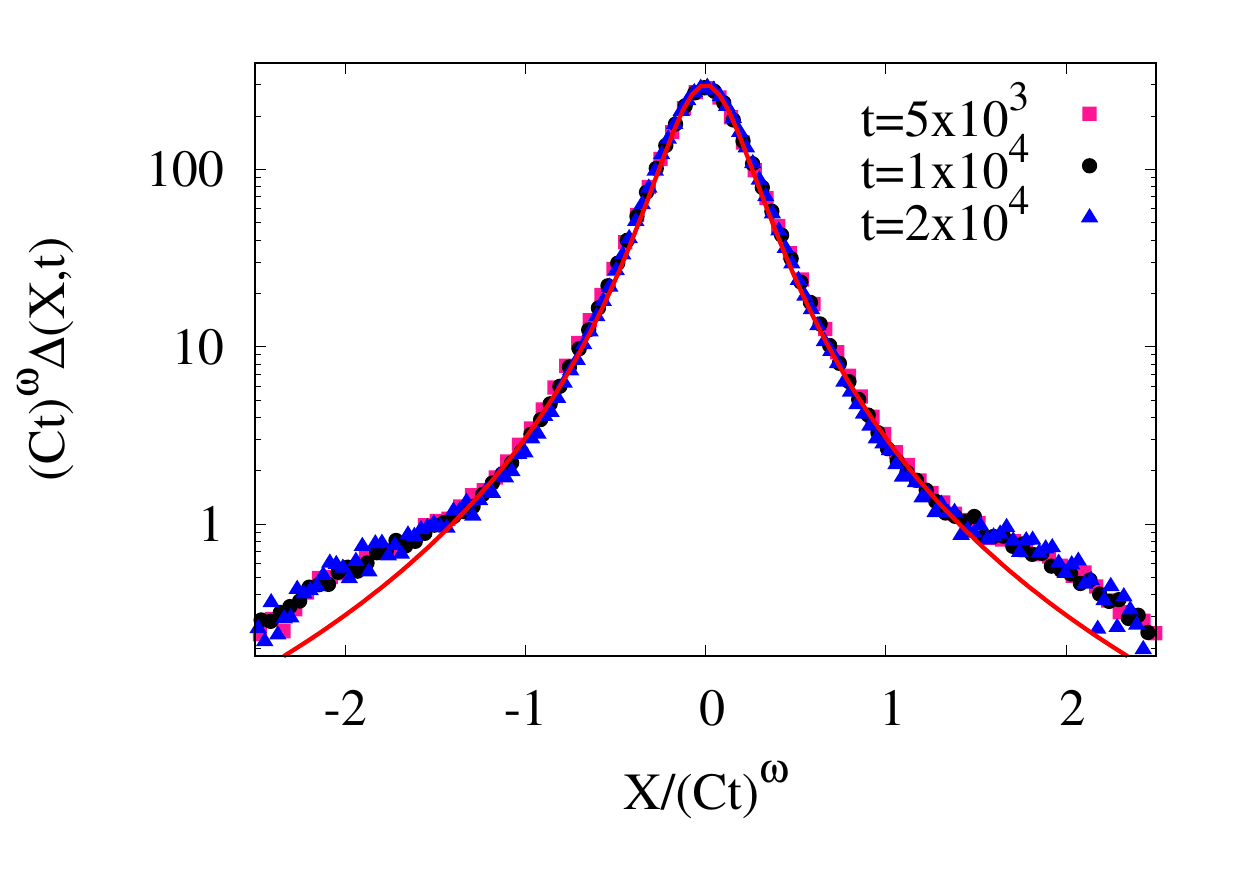}
\caption{{\it Verification of the scaling solution in eq. \eqref{gy}.} Scaled excess density profile $(C t)^{\omega} \Delta(X,t)$, where $\Delta(X,t)=[\rho(X,t)-\rho_c]$, is plotted as a function of the scaling variable $X/(C t)^{\omega}$ for three different times $t =5 \times 10^3$ (pink squares), $10^4$ (black circles) and $2 \times 10^4$ (blue triangles).  Initial density profile is chosen to be the Gaussian one as  given in eq. \eqref{gy-init} with $N_1=150$. Red line - theory [eq. \eqref{gy}] and points - Monte Carlo simulations.}  
\label{fig-gy}
\end{figure}

We now verify in simulations the analytically obtained scaling function in eq. \eqref{gy}. To this end, we consider an initially localized Gaussian density profile,
\begin{equation}
\Delta(X, t=0) = N_1 \frac{1}{\sqrt{2\pi \delta^2}} e^{-X^2/2 \delta^2} ,
\label{gy-init}
\end{equation}
where $\delta=10$ and $N_{1}=150$ are the width and the strength of the profile, respectively.  In simulations, we distribute $N_1=150$ particles, according to the above initial condition, over a critical background configuration of density $\rho_c$. We distribute an appropriate number of particles to generate critical background configurations, having density $\rho_c \simeq 0.94885$ \cite{quasi-periodic}. We run simulations up to three different, but large, times and, in each case, we obtain the excess density profile $\Delta(X, t)$ at the final time by averaging over $2 \times 10^5$ random initial configurations and the corresponding trajectories.  Note that, on large spatio-temporal scales $X \gg \delta$ and $t \gg 1$, the initially localized density profile can be thought of as a Dirac-delta initial condition, which has been used to derive the scaling function  ${\cal G}(y)$ of eq. \eqref{gy}.  In Fig. \ref{fig-gy}, we plot scaled excess density profile ${\cal G}(y) \equiv (C t)^{\omega} \Delta(X, t)$ as a function of the scaling variable $y = X/(C t)^{\omega}$ for three different times $t =5 \times 10^3$ (pink squares), $10^4$ (black circles) and $2 \times 10^4$ (blue triangles). In the same figure, we also compare the simulation results with the analytically obtained scaling function ${\cal G}(y)$ as in eq. \eqref{gy} (red line); here we have used $\beta \simeq 0.42$ and $C \simeq 0.45$, estimated from simulations. We find theory and simulations in quite good agreement over several decades of the scaled excess density values. The deviations at the tails are somewhat expected as the scaling solution eq. \eqref{gy} is not valid in this region due to the break-down of the assumed power-law scaling of activity [eq. \eqref{power-law-a}], which is cut off at very small densities ${\cal O}(\xi^{-1/\nu_{\perp}})$.

\subsubsection{Relaxation of step profile on infinite critical background}

\begin{figure}[H]
\centering
\includegraphics[width=1.0\linewidth]{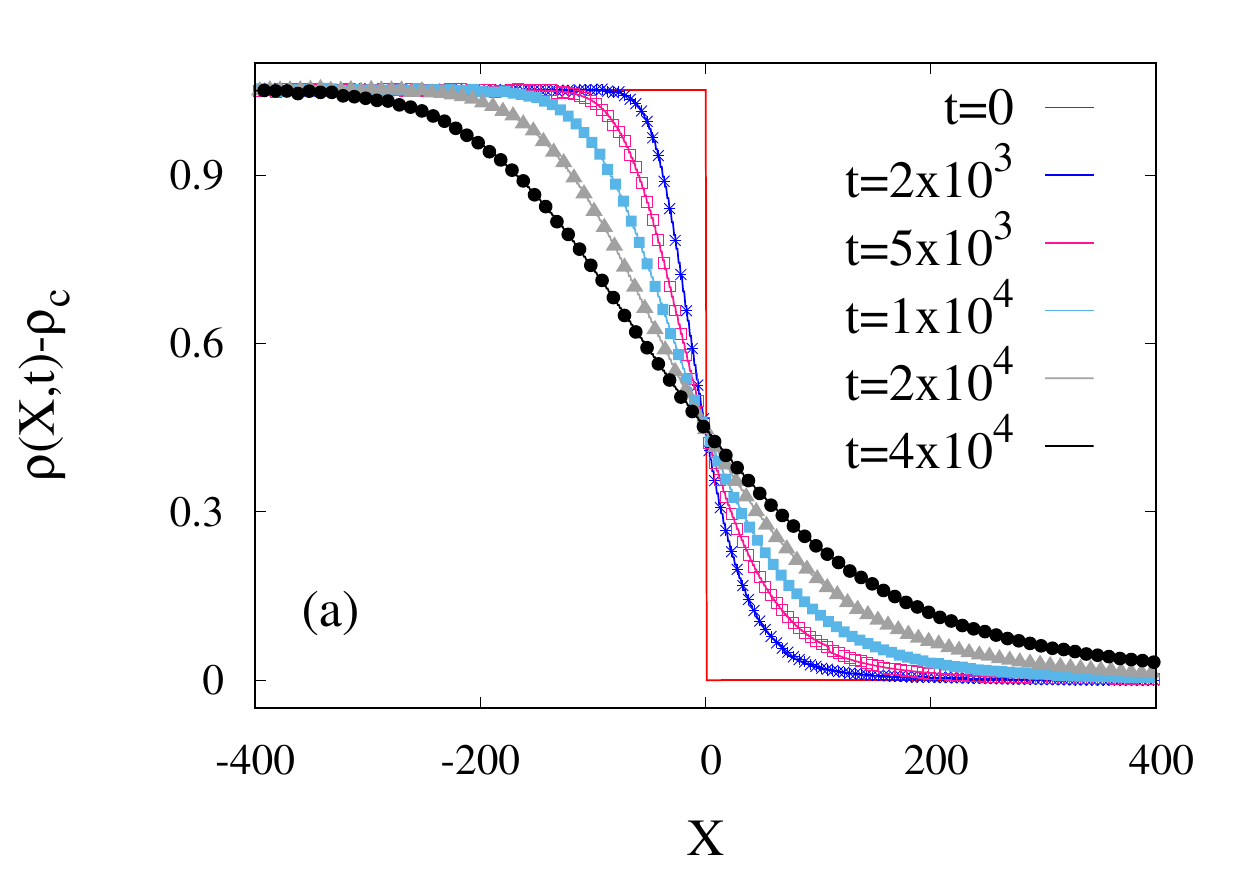}
\includegraphics[width=1.0\linewidth]{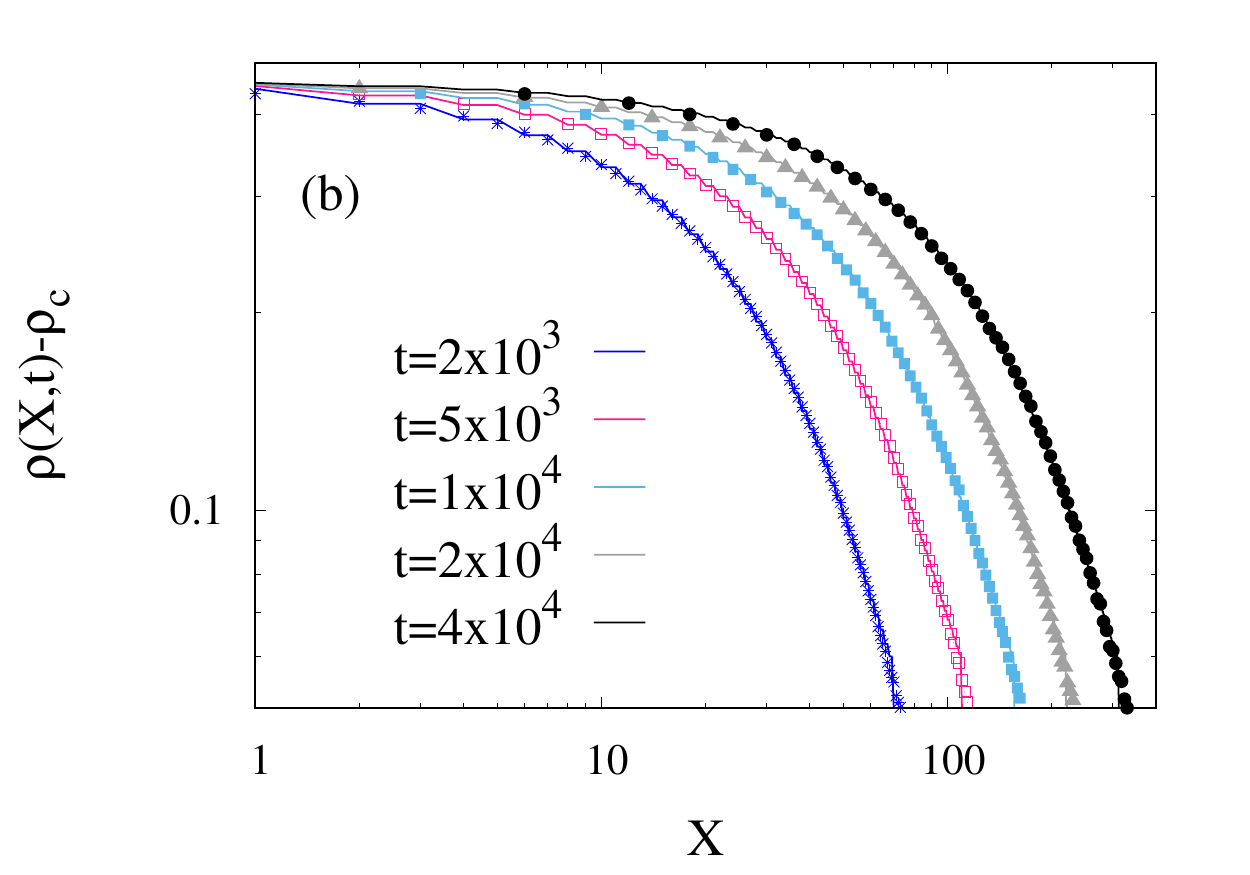}
\includegraphics[width=1.0\linewidth]{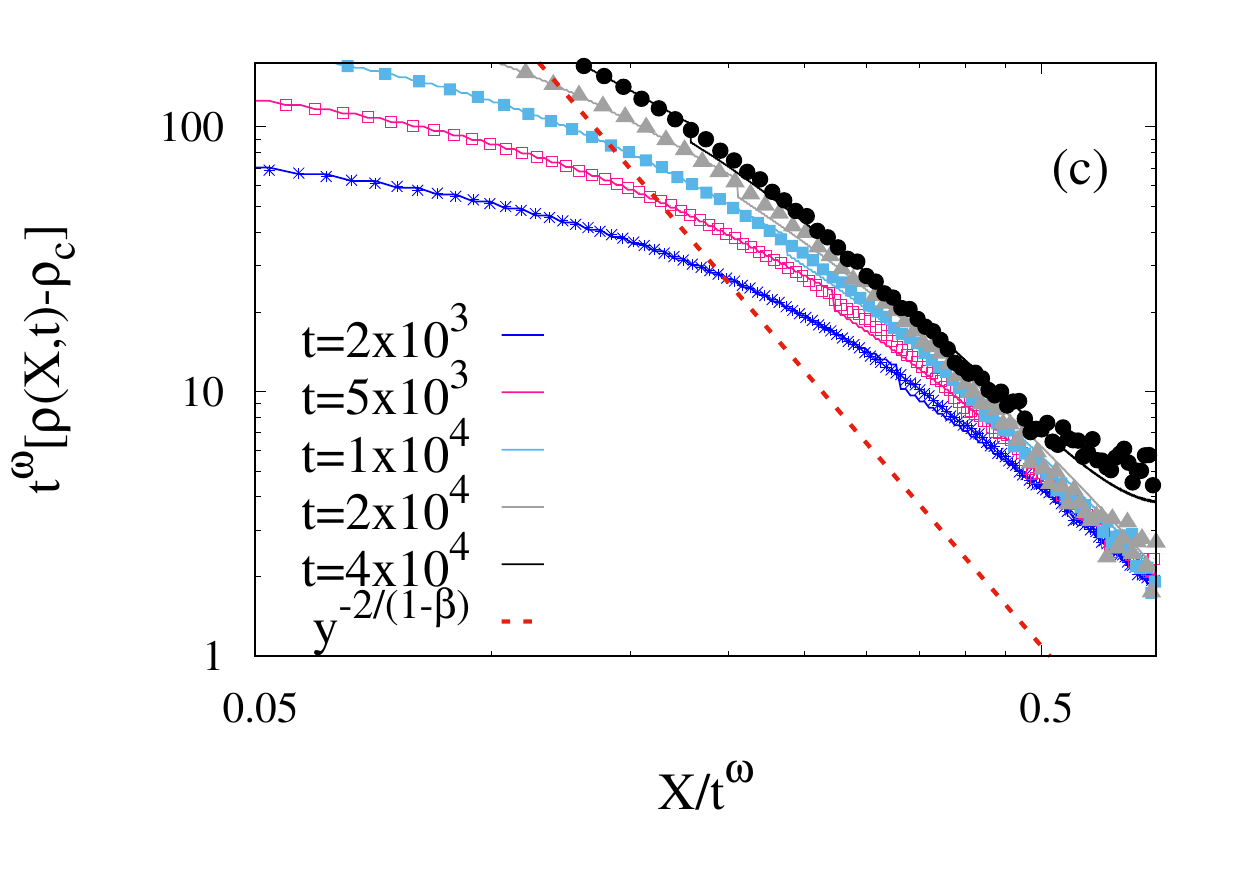}
\caption{  {\it Base density of initial step profile equals to the critical density ($\rho_0 = \rho_c$).} Excess densities, evolved from step initial condition eq. \eqref{step-init}, are plotted as a function of position $X$ at various times $t=2 \times 10^3$ (blue asterisks), $5 \times 10^3$ (pink open squares), $10^4$ (sky-blue filled squares), $2 \times 10^4$ (grey triangles) and $4 \times 10^4$ (black circles). The base density of the step profile is $\rho_0=\rho_c \simeq 0.94885$. Panel (a): The excess density profiles {\it vs.} position are plotted in normal scale. Panel (b): The same as in panel (a) is plotted in log-log scale. Panel (c): The scaled excess density $t^{\omega} \Delta (X, t)$ is plotted as a function of the scaled position $X/t^{\omega}$. Lines, theory [time-integrated eq. \eqref{non-lin-diff}]; points, Monte Carlo simulations.}
\label{equal-rhoc}
\end{figure}

Next we study relaxation of a step-like density profile on infinite critical background. To this end, we consider the initial profile as given in eq. \eqref{step-init} with the base density $\rho_0=\rho_c$ and the height of the profile $\rho_1=1.0$. In panel (a) of Fig. \ref{equal-rhoc}, we plot the excess density $\rho(X,t) - \rho_c$ as a function of position $X$ for various times $t=2 \times 10^3$ (blue asterisks), $5 \times 10^3$ (pink open squares), $10^4$ (sky-blue filled squares), $2 \times 10^4$ (grey triangles) and $4 \times 10^4$ (black circles). In panel (b) of Fig. \ref{equal-rhoc}, we plot the same in log-log scale and observe that the excess density decays with distance from the origin as a power law. Indeed, on large spatio-temporal scales ($X, t \gg 1$), the density field is expected to be described by the scale-invariant solution as given in eq. \eqref{scaling-fn}.  In panel (c) of Fig. \ref{equal-rhoc}, we plot scaled excess density $t^{\omega} [\rho(X, t) - \rho_c]$ as a function of scaled position $X/t^{\omega}$ in a log-log scale, where $\omega=1/(1+\beta)$ [see eq. \eqref{scaling2}] with $\beta \simeq 0.42$. Indeed one could see a reasonable scaling collapse, almost over a decade, where the scaled excess density decays with the scaling variable $y$ as a power law $y^{-2/(1-\beta)}$ [shown by the red guiding dashed line in panel (c)]. However, there is some deviation from scaling at the tails for large times $t \gsim 4 \times 10^4$ as the simulations have been done on large but a finite-size system with periodic boundary condition. From the knowledge of the activity as a function of density, we numerically integrate the nonlinear diffusion equation \eqref{non-lin-diff} to plot the density profiles in panels (a) and (b) of Fig. \ref{equal-rhoc} at the above mentioned times; we find a quite good agreement between our theory (lines) and simulations (points).

\subsection{Density relaxation on sub-critical background}
\label{regime3}

In this section, we consider relaxation of density profiles, where the local density is greater than the critical density in some region, but is less than the critical density elsewhere. We have studied the following three cases. (i) The density relaxation happens on an infinite sub-critical background, where the active region keeps invading the inactive region, with the invasion fronts, separating the active and inactive regions, moving with some velocity. (ii) Next we consider relaxation process on a finite sub-critical background. The global density $\bar{\rho}$ of the initial density profile is taken so that $\bar{\rho} < \rho_{c}$ and the system eventually reaches a frozen state, where the maximum local density in the system is $\rho_c$ in some region and the rest of the system remains uninvaded. (iii) In the third case, we consider relaxation on a finite sub-critical domain, but now the global density $\bar \rho$ of the initial profile is taken so that $\bar{\rho} > \rho_{c}$ and the active region eventually invades the whole system, which becomes super-critical everywhere.

\subsubsection{Relaxation on infinite sub-critical background}

\begin{figure}
\centering
\includegraphics[width=1.0\linewidth]{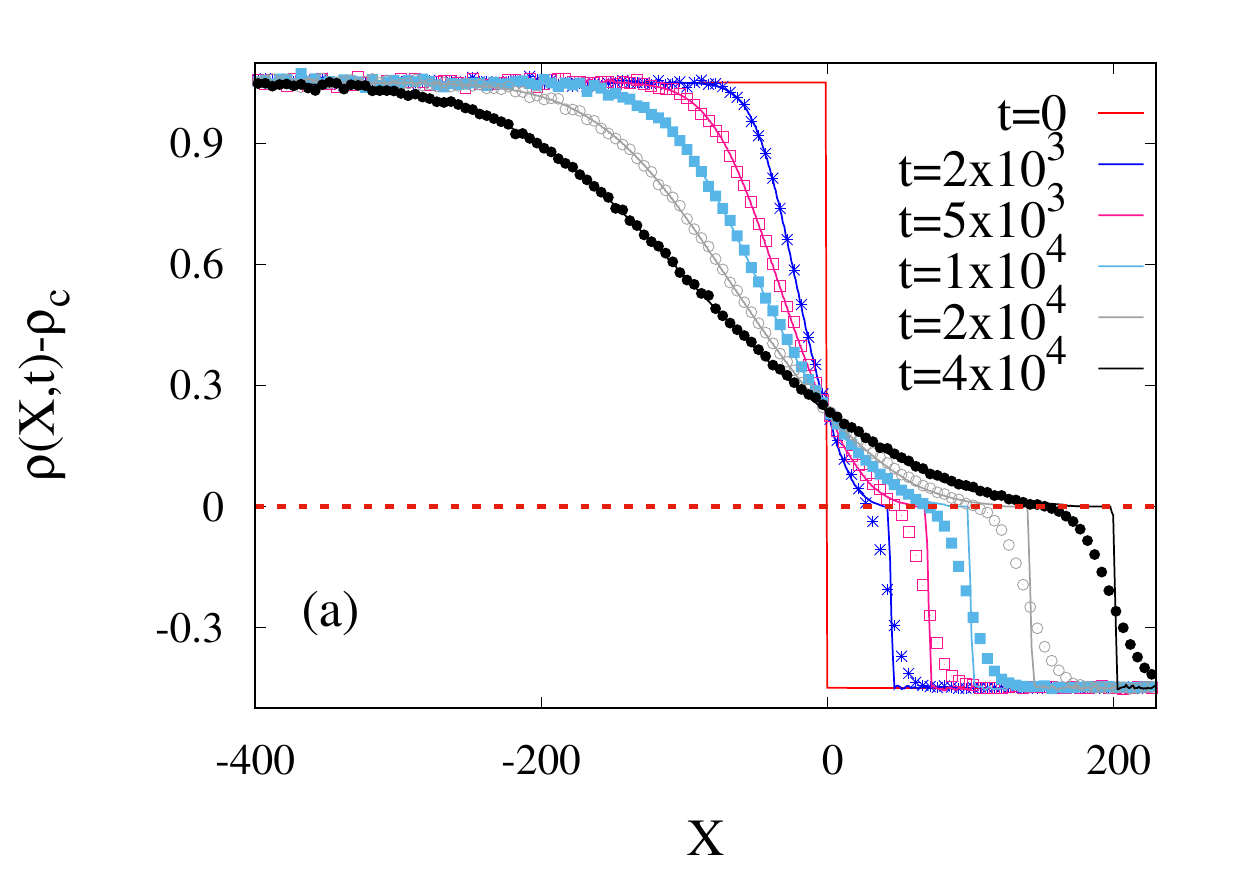}
\includegraphics[width=1.0\linewidth]{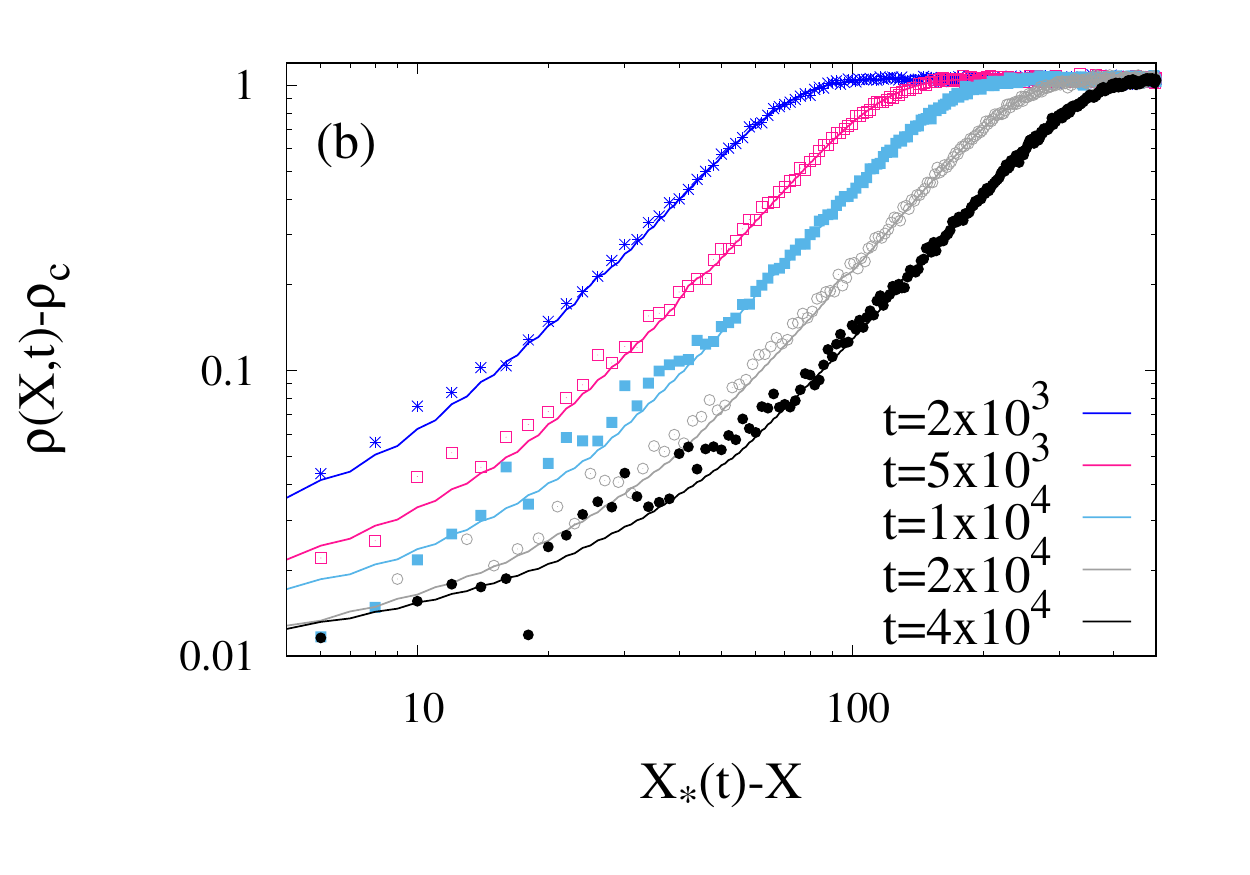}
\includegraphics[width=1.0\linewidth]{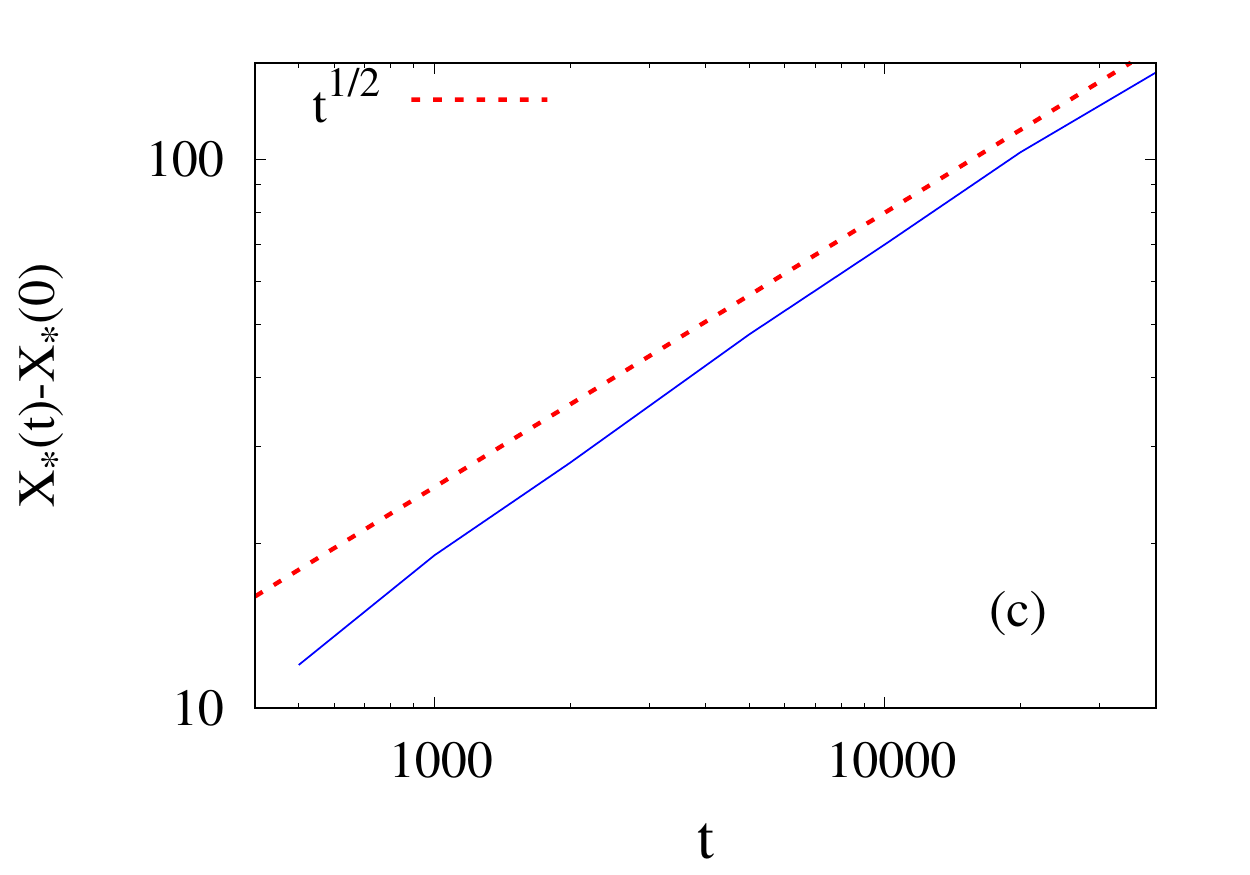}
\caption{ {\it Evolution of density profiles on infinite domain, with base (background) density of the profiles below the critical density ($\rho_0 < \rho_c$).} Excess density profiles $\rho(X,t) - \rho_c$, evolved from the step-like initial profile [eq. \eqref{step-init}], are plotted as a function of position $X$ at various times $t=2 \times 10^3$ (blue asterisks), $5 \times 10^3$ (pink open squares), $10^4$ (sky-blue filled squares), $2 \times 10^4$ (grey open circles) and $4 \times 10^4$ (black filled circles).  Panel (a) - normal scale and panel (b) - log-log scale. Panel (c) - Displacement $X_*(t) - X_*(0)$ of the invasion front is plotted as a function of time $t$. The base, or the background, density of the step profile is $\rho_0=0.5$. Lines - theory [numerically integrated eq. \eqref{non-lin-diff}] and points - simulations.}
\label{below-rhoc}
\end{figure}

First we consider relaxation of an initial step-like profile on an infinite domain, where there are jump discontinuities initially in the local density value at the junctions between active and inactive regions. The step-like density profile at $t=0$ is given by
\begin{eqnarray}
\rho_{in}(X) =\left\lbrace
\begin{array}{ll} 
\rho_{1} + \rho_{0} & \mbox{for} ~ 0 \le X < \delta, \cr
\rho_{0} & \mbox{elsewhere,}           
\end{array}
\right.
\label{step200-900}
\end{eqnarray}
where the height and the width of the step initial profile are $\rho_1$ and $\delta$, respectively; the step profile is generated on a uniform background having density $\rho_0 =0.5 < \rho_c$ (the same base density is considered in the rest of the paper). In simulations, we take $L, \delta \gg 1$ to study the density relaxation in an infinite domain. In panel (a) of Fig. \ref{below-rhoc}, we plot the excess density $\rho(X, t) -\rho_c$, measured around the critical density,  as a function of position $X$ at various times $t=2 \times 10^3$ (blue asterisks), $5 \times 10^3$ (pink open squares), $10^4$ (sky-blue filled squares), $2 \times 10^4$ (grey open circles) and $4 \times 10^4$ (black filled circles).  At later times $t>0$, the density profile develops a boundary layer, which decays over a finite length scale from the critical background density to the base (background) density. The boundary layer can be characterized by an invasion front, located at position $X_*(t)$ at time $t$, which moves forward with a time-dependent velocity. Here the invasion front is operationally defined to be the position where the density profile falls off to the value $\rho(X=X_*, t)=\rho_0+(\rho_c-\rho_0)/2$ (i.e., the midpoint between the critical density and the base, or the background, density). In panel (b) of Fig. \ref{below-rhoc}, we plot the excess density $\rho(X,t) - \rho_c$ as a function of a shifted position variable $\delta X = X_*(t) - X$, which is measured from the position of the propagating front. One can see that the density profile $\rho(X, t)-\rho_c \sim (\delta X)^{\gamma}$ grows with the shifted position $\delta X$ as a power law, where the exponent is estimated to be $\gamma \simeq 1.5$ from simulations.  In panel (c) of Fig. \ref{below-rhoc}, we plot the position of the invasion front $X_*(t)$ as a function of time $t$ and find that the front moves with a time-dependent velocity $v_*(t) = X_*(t)/t$. In fact, the position of the density front $X_*(t) \sim t^{\alpha}$ grows sub-linearly with time $t$ where the exponent $\alpha \simeq 1/2$ [shown by the red dashed guiding line in panel (c) of Fig. \ref{below-rhoc}].

To compare the above simulation results with the theory,  we numerically integrate the nonlinear diffusion equation \eqref{non-lin-diff}, which should describe the large scale spatio-temporal evolution of the density field. Now, to integrate eq. \eqref{non-lin-diff}, we require explicit functional dependence of activity $a(\rho)$ on density $\rho$, which is determined from simulations; note that, for sub-critical density $\rho < \rho_c$, we use $a(\rho)=0$ in eq. \eqref{non-lin-diff}, which is the case in the inactive regions (absorbing phase), where the bulk-diffusion coefficient vanishes. In Fig. \ref{below-rhoc}, one can see that the numerically integrated density profile (lines) obtained by integrating eq. \eqref{non-lin-diff}, with the above functional form of the activity, not only captures quite nicely the simulation results (points) almost over a couple of decades of the density values, but also predicts the position of the moving density front quite precisely. Note that the boundary layers around the invasion fronts get smeared and their exact functional forms are not captured by our theory, which predicts a sharp discontinuous jump in densities at the front positions.
However, as the width of the boundary layer should not increase with system size $L$, we expect to recover the discontinuous jump in density in the limit of large $L$. The above results remain qualitatively the same also for a wedge-like profile, which is not presented here.

\subsubsection{Relaxation to frozen state: Global density $\bar{\rho} < \rho_{c}$.}

 Next we consider density relaxation on a finite domain. We consider two kinds of  initial density profiles: a step-like profile as in eq. \eqref{step200-900} with a finite width $\delta$ and a wedge-like density profile given by
\begin{eqnarray}
\rho_{in}(X) =\left\lbrace
\begin{array}{ll} 
\rho_{0} + \rho_{1}(X - X_1)/\delta & \mbox{for} ~ X_1 \leq X < L/2, \cr
\rho_{0} + \rho_{1}(X_2 - X)/\delta & \mbox{for} ~ L/2\leq X < X_2, \cr
\rho_{0} & \mbox{otherwise.}           
\end{array}
\right.
\label{wedge200-900}
\end{eqnarray}
The wedge-like initial profile is centered at $X=L/2$, has a width $\delta$ and height $\rho_1$; we set $X_1=(L-\delta)/2$ and $X_2=(L+\delta)/2$.  The wedge profile is generated over a uniform background having density $\rho_0 < \rho_c$. We fix the width $\delta = 200$ for both the profiles, and the height $\rho_1 = 1.5$ for the step profile and $\rho_1 = 3.5$ for the wedge profile. We set the global density $\bar{\rho} = 0.8 < \rho_{c}$ below the critical density so that the system eventually evolves to a frozen (absorbing) state. We generate initial density profiles by randomly distributing $N_1 = L(\bar{\rho} - \rho_0) = 300$ number of particles over a uniform background, having density $\rho_0$ and over a domain $-\delta/2 < X < \delta/2$.

\begin{figure}[H]
\centering
\includegraphics[width=1.0\linewidth]{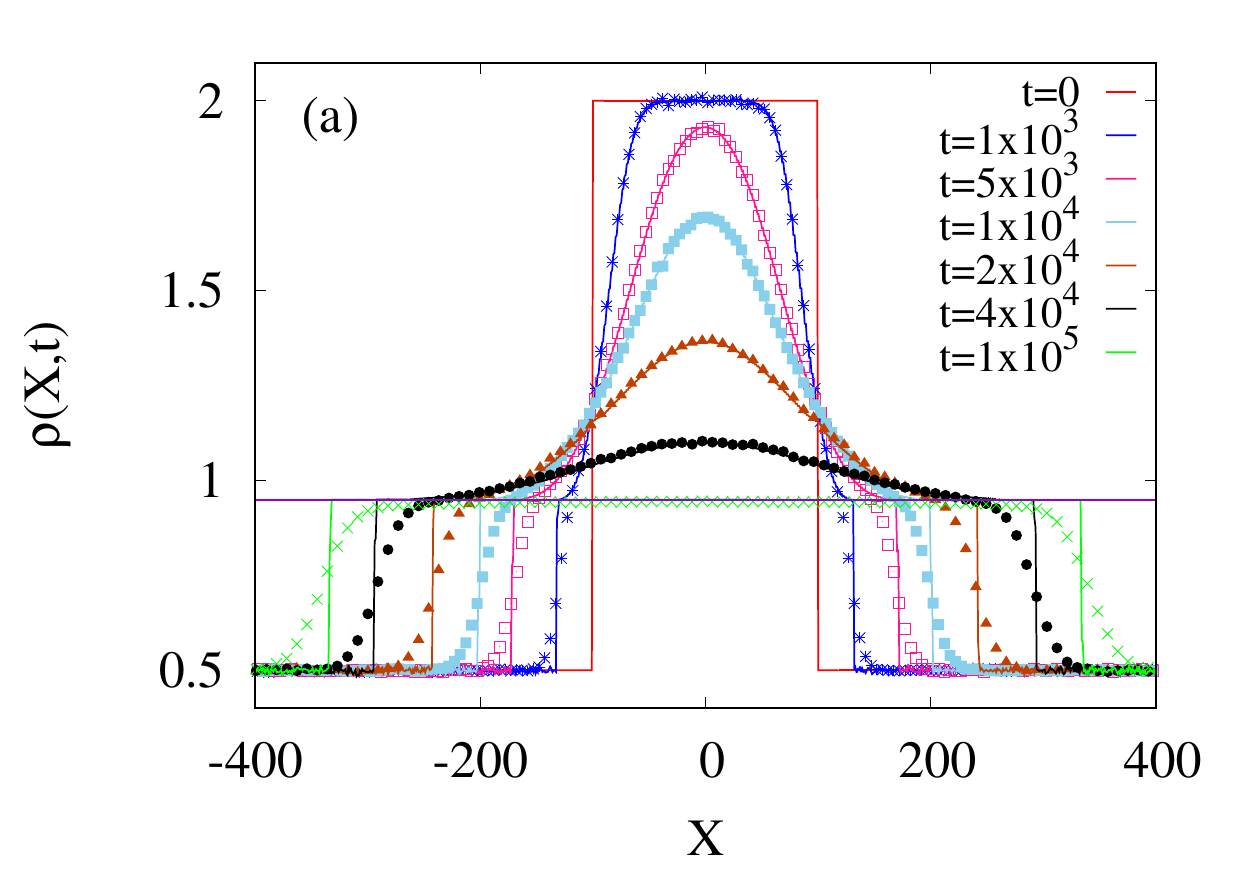}
\includegraphics[width=1.0\linewidth]{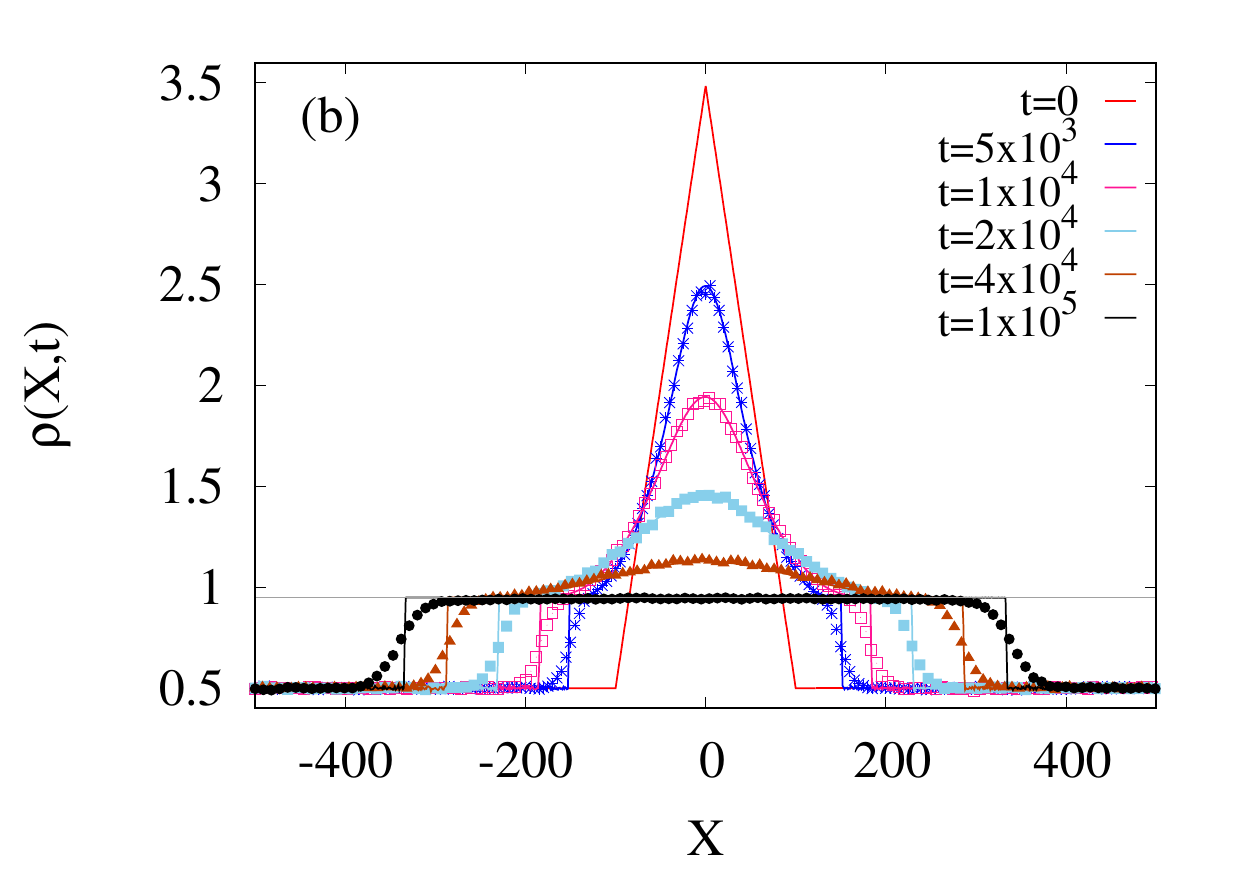}
\caption{ {\it Relaxation to a fully absorbing state, from initial profiles made of active and inactive regions.} Time-dependent density profiles are plotted for (a) step and (b) wedge initial profiles, having  a uniform base (background) density $\rho_{0} = 0.5 < \rho_{c}$ and height $\rho_1=2.0$ (step) and $3.5$ (wedge). Panel (a): Density profiles are plotted at  times $t=10^3$ (blue asterisks), $5 \times 10^3$ (pink open squares), $10^4$ (sky-blue filled squares), $2 \times 10^4$ (brown filled triangles), $4 \times 10^4$ (black filled circles) and $10^5$ (green crosses). Panel (b): Density profiles are plotted at   times $t = 5 \times 10^3$ (blue asterisks), $10^4$ (pink open squares), $2 \times 10^4$ (sky-blue filled squares), $4 \times 10^4$ (brown filled triangles) and $10^5$ (black filled circles). Global density $\bar \rho = 0.8 < \rho_c$; points - Monte Carlo simulations; lines - theory.}
\label{width200-N300}
\end{figure} 

In panel (a) of Fig. \ref{width200-N300}, we plot the density profile $\rho(X, t)$ as a function of position $X$ at various times $t=10^3$ (blue asterisks), $5 \times 10^3$ (pink open squares), $10^4$ (sky-blue filled squares), $2 \times 10^4$ (brown filled triangles), $4 \times 10^4$ (black filled circles) and $10^5$ (green crosses). In panel (b) of Fig. \ref{width200-N300}, the density profiles are plotted at times $t = 5 \times 10^3$ (blue asterisks), $10^4$ (pink open squares), $2 \times 10^4$ (sky-blue filled squares), $4 \times 10^4$ (brown filled triangles) and $10^5$ (black filled circles). The grey horizontal line denotes the critical density $\rho_{c} \simeq 0.94885$. 
From the above plots, it is observed that, irrespective of the shapes of the initial profiles, the time-evolved density profiles eventually become equal to the critical density $\rho_{c}$ and gets frozen as the systems move into an absorbing state everywhere.  In Fig. \ref{width200-N300}, we have also compared the density profiles obtained from simulations with that obtained by numerically integrating the nonlinear diffusion equation \eqref{non-lin-diff}. 
According to the theory, we expect that the density at long times should be equal to the critical density in any region that initially started active, or got invaded to become active. On the other hand, in the uninvaded region, the density remains equal to the initial density. We do not exactly see this behavior in simulations as the size of the invaded regions fluctuate from sample to sample and are actually ranging over a finite region of space, thus smearing out the theoretically predicted discontinuity in simulations. However, the smeared invasion front gets arbitrarily sharp on the macroscopic scales, upon increasing the initially added particle number.

\subsubsection{Relaxation from sub-critical to super-critical state: Global density $\bar{\rho} > \rho_c$.}

\begin{figure}[H]
\centering
\includegraphics[width=1.0\linewidth]{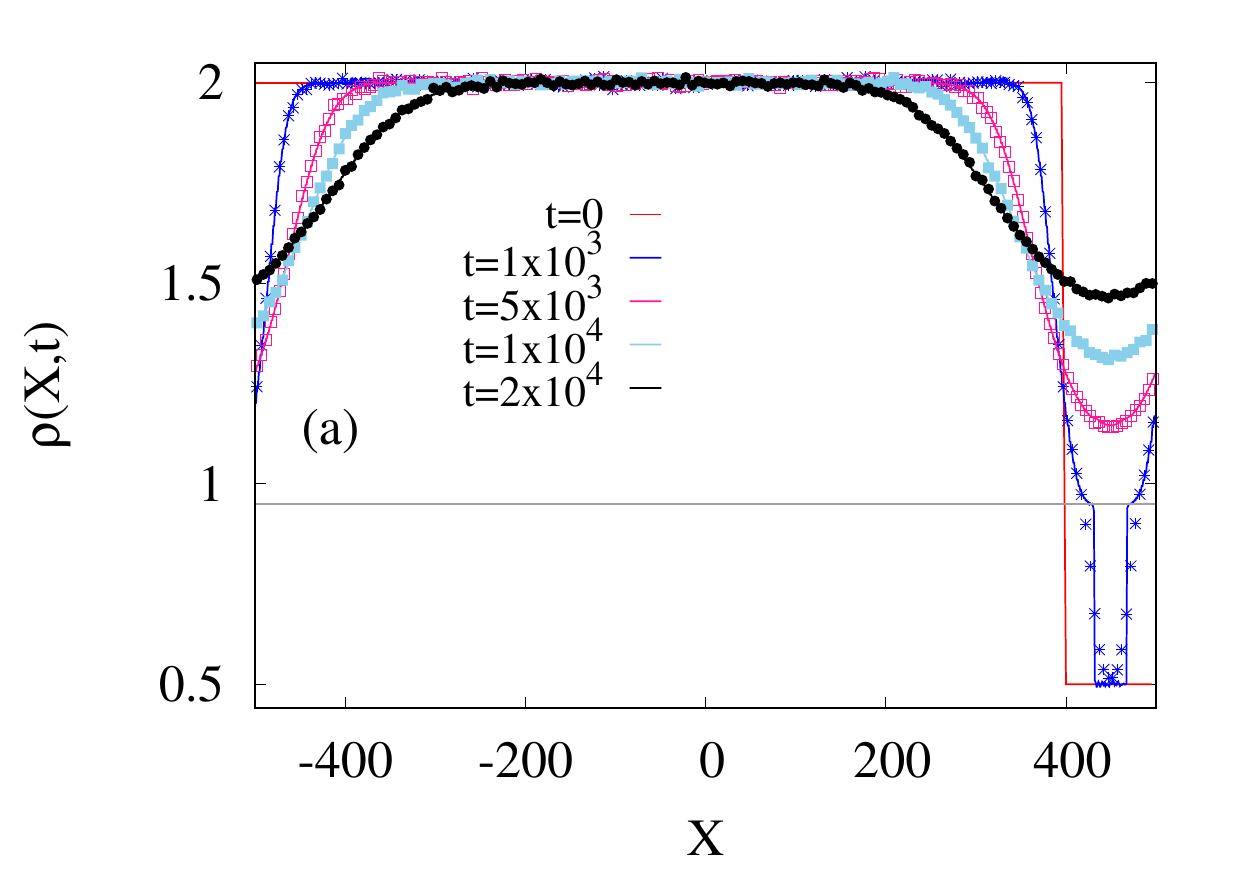}
\includegraphics[width=1.0\linewidth]{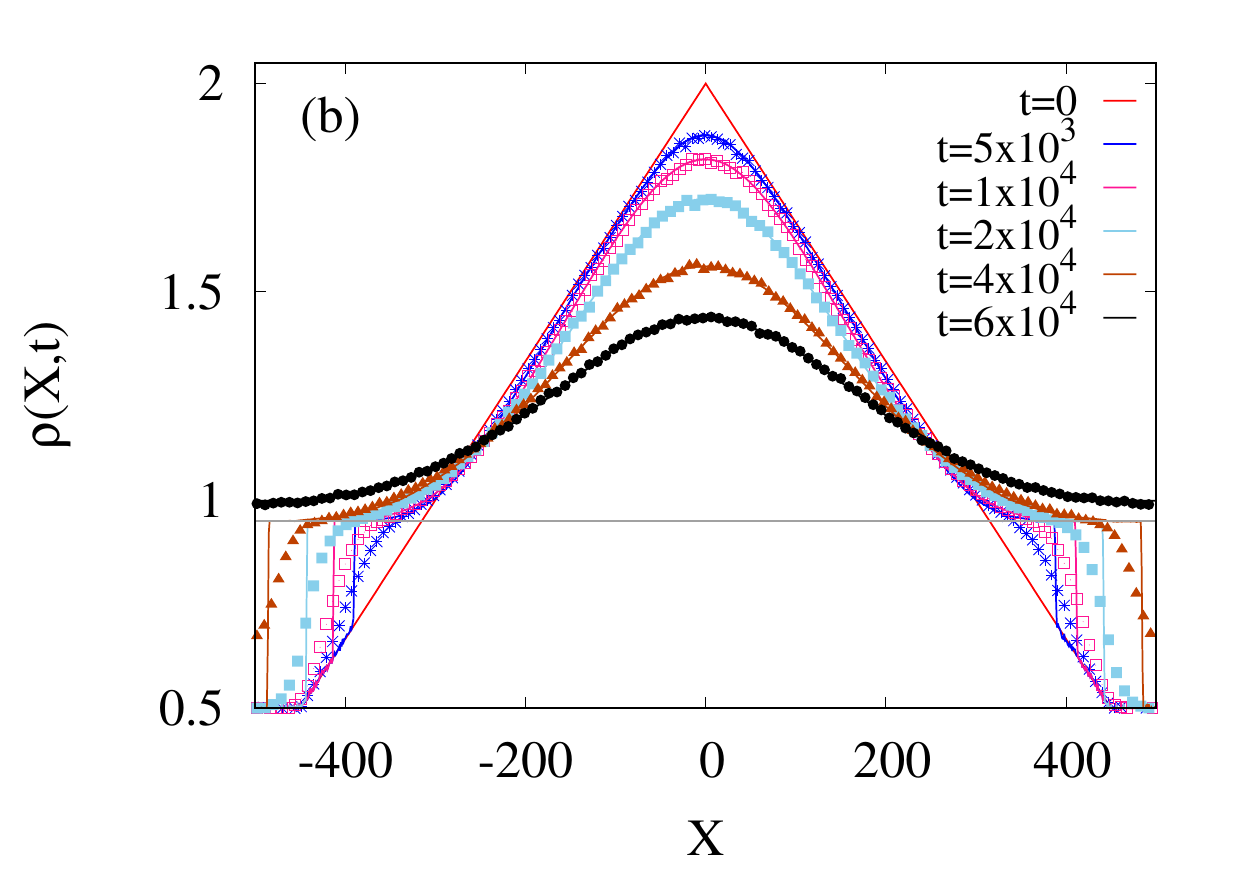}
\caption{ {\it Relaxation to a fully super-critical state, from initial profiles made of active and inactive regions.}  Time-dependent profiles are plotted for (a) step and (b) wedge initial profiles, having a uniform base or background density $\rho_0 = 0.5 < \rho_c$ and height $\rho_1 = 2.0$. Panel (a): Density profiles are plotted at  times $t=10^3$ (blue asterisks), $5 \times 10^3$ (pink open squares), $10^4$ (sky-blue filled squares) and $2 \times 10^4$ (black filled circles);  global density is $\bar{\rho} = 1.85 > \rho_c$. Panel (b): Density profiles are plotted at  times $t=10^3$ (blue asterisks), $5 \times 10^3$ (pink open squares), $10^4$ (sky-blue filled squares), $2 \times 10^4$ (brown filled triangles), $4 \times 10^4$ (black filled circles) and $6 \times 10^4$ (green crosses);  global density is $\bar{\rho} = 1.175 > \rho_c$.  Points - Monte Carlo simulations; lines - theory.}
\label{width900}
\end{figure}

Finally we consider relaxation of initial  density profiles where the global density $\bar \rho$ is chosen such that $\bar \rho > \rho_c$. Therefore, in this case, the system finally evolves to an active or super-critical state everywhere. 
The width of the initial density profiles is taken to be $\delta = 900$ and the initial piles are formed over a uniform background having density $\rho_0 = 0.5$ (which is much below the critical density).   In panel (a) of Fig. \ref{width900}, we plot the density profile $\rho(X, t)$ as a function of position $X$ at various times $t=10^3$ (blue asterisks), $5 \times 10^3$ (pink open squares), $10^4$ (sky-blue filled squares) and $2 \times 10^4$ (black filled circles) for step initial profile. Whereas, in panel (b) of the same figure, we plot the density profiles at times $t= 5 \times 10^3$ (blue asterisks), $10^4$ (pink open squares) and $2 \times 10^4$ (sky-blue filled squares) $4 \times 10^4$ (brown filled triangles) and $6 \times 10^4$ (black filled circles). The grey horizontal line denotes the critical density $\rho_{c}$. In the case of step-like initial profile, $N_{1} = L(\bar \rho - \rho_0) = 1350$ particles and, in the case of wedge-like initial profile, $N_1 = L(\bar \rho - \rho_0) = 675$ numbers of particles are distributed over an uniform background density $\rho_0$, keeping the height of the step and the peak of the wedge at $\rho_1 = 1.5$; we take $L=1000$. The global density $\bar \rho$ is kept at $1.85$ for the step profile and $1.175$ for the wedge-like profile. 

From the above plots, we observe that, for both the initial profiles, the regions of the density profiles, which were below critical density $\rho_{c}$ and therefore were in the frozen state initially, gradually become active due to the  invasion of the active regions into the inactive ones. Eventually, the inactive regions are lifted above the critical density and the systems become active everywhere. This happens because the global density $\bar \rho$ is chosen to be greater than the critical density $\rho_c \simeq 0.94885$. We find that, for the step-like profile, the inactive region becomes active at a comparatively smaller time ($\sim 2 \times 10^4$) than the time ($\sim 6 \times 10^4$) required for the wedge-like initial profile. We have compared the above density profiles obtained from simulations with that obtained by numerically integrating the nonlinear diffusion equation \eqref{non-lin-diff}. We find  an excellent agreement between the theory (lines) and simulations (points), except at the boundary layer regions around the invasion fronts, where the density discontinuities predicted by our theory get smeared in the actual simulations.

\section{Summary and concluding remarks}
\label{sec-summary}

In this paper, we study density relaxation in the Manna sandpile with conserved mass and continuous-time dynamics. We recently proposed a theory for large-scale (hydrodynamic) time evolution of density perturbations in conserved stochastic sandpiles. Using these ideas and through direct Monte Carlo simulations, here we investigate in detail the density relaxations in different situations in the conserved Manna sandpiles in Ref. \cite{Chatterjee_PRE2018}. We consider density perturbations typically having small wave numbers $k \rightarrow 0$ and relaxing on finite (periodic) as well as infinite domains. 

Far from criticality, where the correlation length $\xi$ is finite [i.e., in the density regime $\Delta(X, t) =\rho(X, t)-\rho_c \gsim 1$], relaxation of long-wavelength density perturbations in the limit of $k \xi \ll 1$ is diffusive in nature. Indeed the time evolution of  initial density profiles in this case is governed by a nonlinear diffusion equation \eqref{diffusion_continuum}, with a density-dependent bulk-diffusion coefficient $D(\rho) = d a(\rho)/d \rho$, where $a(\rho)$ is the steady-state activity and is a nonlinear function of coarse-grained density $\rho$. Consequently, the relaxation times for the long-wavelength density perturbations vary as  $\tau_R \sim k^{-2}/D(\rho)$, i.e., in a system of size $L$,  relaxation time $\tau_R \sim L^2/D(\rho)$.

Near criticality where correlation length $\xi \gg 1$ large and $k \xi \gsim 1$ (i.e., in the density regime $\Delta(X, t) \sim \xi^{-1/\nu_{\perp}} \ll 1$), the diffusive scaling breaks down as the bulk-diffusion coefficient diverges as $D \sim \xi^{(1-\beta)/\nu_{\perp}}$. Consequently, particle transport becomes  anomalous, leading to the relaxation times for long-wavelength density perturbations  vary as $\tau_R \sim k^{-z}$, where the dynamic exponent $z=2 - (1-\beta)/\nu_{\perp} < 2$ is determined in terms of the near-critical order-parameter exponent $\beta$ and correlation-length exponent $\nu_{\perp}$. 
Note that the relaxation time at the critical point is still infinite for infinite systems as the
critical exponent $z$ is defined in the limit of wave number $k$ being small, by keeping $\Delta$ fixed at a very small value. Interestingly, for a small fixed $k$, the relaxation time is smaller
than that away from criticality; that is, the relaxation time decreases as $\Delta$ decreases.

In the density regime where $\xi \gg 1$, but $k \xi \ll 1$ (i.e., when $L^{-1/\nu_{\perp}} \ll \Delta(X,t) \lsim 1$), relaxation of an initially localized density perturbation on infinite critical background exhibits a self-similar structure on large space and time scales $X, t \gg 1$. In this case, we exactly determine, within our hydrodynamic theory [see eq. \eqref{diffusion_equation_putting_activity}], the asymptotic scaling form ${\cal G}(y) \propto (B_0+ B y^2)^{-b}$ [see eq. \eqref{gy}] for the scaled time-dependent (excess) density profile $t^{-\omega} \Delta(X,t)$ as a function of the single scaling variable $y \sim X/t^{\omega}$, with the exponents $\omega=1/(1+\beta) >1/2$ and $b = 1/(1-\beta)$; here $B_0$ and $B$ are two integration constants, which can be determined from the normalization condition and in terms of the exponent $\beta$.

We have also studied the cases where local density can be less than the critical density in some regions and greater than the critical density elsewhere. The active region invades into the inactive regions and, eventually, the system goes to either a frozen (absorbing) state ($\bar \rho < \rho_c$) or a super-critical state ($\bar \rho > \rho_c$), depending on the global density $\bar \rho$. 
In these cases, the predictions of our hydrodynamic theory are in an excellent agreement with simulations, except at the regions around the invasion fronts, where the front of the invaded region  lead to a smeared profile, while the theory predicts a discontinuity in the infinite L limit. 
However, the jump in the density will get sharper in the reduced variable $x= X/L$
as system size $L$ is increased, and we would expect to recover the discontinuous jump in the limit of $L$ being large.

We believe that these studies of density relaxation in conserved Manna sandpiles more generally would provide some useful insights into the exact hydrodynamic structure of sandpiles. Indeed, the hydrodynamic theory developed here could open up an exciting avenue for characterizing fluctuations not only in the conserved version, but also in the ``self-organized critical'' (SOC), or the driven dissipative, version of sandpiles.

\section{Acknowledgment}

We acknowledge the hospitality at the International Centre for Theoretical Sciences (ICTS), Bengaluru during a visit for participating in the program ``Universality in random structures: Interfaces, Matrices, Sandpiles'' (Code: ICTS/URS2019/01), where some of the initial ideas about the project were conceived. P.P. acknowledges the Science and Engineering Research Board (SERB), India, under Grant No. MTR/2019/000386, for financial support. 

\bibliography{bib_manna_hydrodynamics1}

\begin{thebibliography}{45}
\expandafter\ifx\csname natexlab\endcsname\relax\def\natexlab#1{#1}\fi
\expandafter\ifx\csname bibnamefont\endcsname\relax
  \def\bibnamefont#1{#1}\fi
\expandafter\ifx\csname bibfnamefont\endcsname\relax
  \def\bibfnamefont#1{#1}\fi
\expandafter\ifx\csname citenamefont\endcsname\relax
  \def\citenamefont#1{#1}\fi
\expandafter\ifx\csname url\endcsname\relax
  \def\url#1{\texttt{#1}}\fi
\expandafter\ifx\csname urlprefix\endcsname\relax\def\urlprefix{URL }\fi
\providecommand{\bibinfo}[2]{#2}
\providecommand{\eprint}[2][]{\url{#2}}

\bibitem[{\citenamefont{Bak et~al.}(1987)\citenamefont{Bak, Tang, and
  Wiesenfeld}}]{Bak_PRL1987}
\bibinfo{author}{\bibfnamefont{P.}~\bibnamefont{Bak}},
  \bibinfo{author}{\bibfnamefont{C.}~\bibnamefont{Tang}}, \bibnamefont{and}
  \bibinfo{author}{\bibfnamefont{K.}~\bibnamefont{Wiesenfeld}},
  \bibinfo{journal}{Phys. Rev. Lett.} \textbf{\bibinfo{volume}{59}},
  \bibinfo{pages}{381} (\bibinfo{year}{1987}),
  \urlprefix\url{https://link.aps.org/doi/10.1103/PhysRevLett.59.381}.

\bibitem[{\citenamefont{Kirkby}(1983)}]{Mountain}
\bibinfo{author}{\bibfnamefont{M.~J.} \bibnamefont{Kirkby}},
  \bibinfo{journal}{Earth Surface Processes and Landforms}
  \textbf{\bibinfo{volume}{8}}, \bibinfo{pages}{406} (\bibinfo{year}{1983}),
  \urlprefix\url{https://doi.org/10.1002/esp.3290080415}.

\bibitem[{\citenamefont{Scheidegger}(1967)}]{river}
\bibinfo{author}{\bibfnamefont{A.~E.} \bibnamefont{Scheidegger}},
  \bibinfo{journal}{International Association of Scientific Hydrology.
  Bulletin} \textbf{\bibinfo{volume}{12}}, \bibinfo{pages}{57}
  (\bibinfo{year}{1967}), \eprint{https://doi.org/10.1080/02626666709493550},
  \urlprefix\url{https://doi.org/10.1080/02626666709493550}.

\bibitem[{\citenamefont{Dutta and Horn}(1981)}]{metal}
\bibinfo{author}{\bibfnamefont{P.}~\bibnamefont{Dutta}} \bibnamefont{and}
  \bibinfo{author}{\bibfnamefont{P.~M.} \bibnamefont{Horn}},
  \bibinfo{journal}{Rev. Mod. Phys.} \textbf{\bibinfo{volume}{53}},
  \bibinfo{pages}{497} (\bibinfo{year}{1981}),
  \urlprefix\url{https://link.aps.org/doi/10.1103/RevModPhys.53.497}.

\bibitem[{\citenamefont{Bak and Tang}(1989)}]{Earthquake}
\bibinfo{author}{\bibfnamefont{P.}~\bibnamefont{Bak}} \bibnamefont{and}
  \bibinfo{author}{\bibfnamefont{C.}~\bibnamefont{Tang}},
  \bibinfo{journal}{Journal of Geophysical Research: Solid Earth}
  \textbf{\bibinfo{volume}{94}}, \bibinfo{pages}{15635} (\bibinfo{year}{1989}),
  \urlprefix\url{https://doi.org/10.1029/JB094iB11p15635}.

\bibitem[{\citenamefont{Andrade et~al.}(1998)\citenamefont{Andrade,
  Schellnhuber, and Claussen}}]{Rain}
\bibinfo{author}{\bibfnamefont{R.}~\bibnamefont{Andrade}},
  \bibinfo{author}{\bibfnamefont{H.}~\bibnamefont{Schellnhuber}},
  \bibnamefont{and} \bibinfo{author}{\bibfnamefont{M.}~\bibnamefont{Claussen}},
  \bibinfo{journal}{Physica A: Statistical Mechanics and its Applications}
  \textbf{\bibinfo{volume}{254}}, \bibinfo{pages}{557 } (\bibinfo{year}{1998}),
  ISSN \bibinfo{issn}{0378-4371},
  \urlprefix\url{http://www.sciencedirect.com/science/article/pii/S0378437198000570}.

\bibitem[{\citenamefont{{R. Chialvo}}(2004)}]{Brain}
\bibinfo{author}{\bibfnamefont{D.}~\bibnamefont{{R. Chialvo}}},
  \bibinfo{journal}{Physica A: Statistical Mechanics and its Applications}
  \textbf{\bibinfo{volume}{340}}, \bibinfo{pages}{756 } (\bibinfo{year}{2004}),
  ISSN \bibinfo{issn}{0378-4371}, \bibinfo{note}{complexity and Criticality: in
  memory of Per Bak (1947--2002)},
  \urlprefix\url{http://www.sciencedirect.com/science/article/pii/S0378437104005734}.

\bibitem[{\citenamefont{Sethna et~al.}(2001)\citenamefont{Sethna, Dahmen, and
  Myers}}]{Sethna_Nat2001}
\bibinfo{author}{\bibfnamefont{J.~P.} \bibnamefont{Sethna}},
  \bibinfo{author}{\bibfnamefont{K.~A.} \bibnamefont{Dahmen}},
  \bibnamefont{and} \bibinfo{author}{\bibfnamefont{C.~R.} \bibnamefont{Myers}},
  \bibinfo{journal}{Nature} \textbf{\bibinfo{volume}{410}},
  \bibinfo{pages}{242–250} (\bibinfo{year}{2001}),
  \urlprefix\url{https://doi.org/10.1038/35065675}.

\bibitem[{\citenamefont{Peters and Neelin}(2006)}]{Peters_Nature2006}
\bibinfo{author}{\bibfnamefont{O.}~\bibnamefont{Peters}} \bibnamefont{and}
  \bibinfo{author}{\bibfnamefont{J.~D.} \bibnamefont{Neelin}},
  \bibinfo{journal}{Nature Physics} \textbf{\bibinfo{volume}{2}},
  \bibinfo{pages}{393} (\bibinfo{year}{2006}), ISSN \bibinfo{issn}{1745-2481},
  \urlprefix\url{https://doi.org/10.1038/nphys314}.

\bibitem[{\citenamefont{Jensen}(1998)}]{book-Jensen}
\bibinfo{author}{\bibfnamefont{P.~H.~J.} \bibnamefont{Jensen}}, Self-Organized
  Criticality: Emergent Complex Behavior in Physical and Biological Systems
  (\bibinfo{publisher}{Cambridge University Press}, \bibinfo{year}{1998}), ISBN
  \bibinfo{isbn}{0521483719,9780521483711},
  \urlprefix\url{http://gen.lib.rus.ec/book/index.php?md5=faf3e690197e6c85aa826a4ec440cf75}.

\bibitem[{\citenamefont{{Aschwanden}}(2013)}]{Aschwanden-2013}
\bibinfo{author}{\bibfnamefont{M.~J.} \bibnamefont{{Aschwanden}}},
  \emph{\bibinfo{title}{{Self-Organized Criticality Systems}}}
  (\bibinfo{publisher}{Open Academic Press, Berlin}, \bibinfo{year}{2013}).

\bibitem[{\citenamefont{Watkins et~al.}(2016)\citenamefont{Watkins, Pruessner,
  Chapman, Crosby, and Jensen}}]{Watkins-2016}
\bibinfo{author}{\bibfnamefont{N.~W.} \bibnamefont{Watkins}},
  \bibinfo{author}{\bibfnamefont{G.}~\bibnamefont{Pruessner}},
  \bibinfo{author}{\bibfnamefont{S.~C.} \bibnamefont{Chapman}},
  \bibinfo{author}{\bibfnamefont{N.~B.} \bibnamefont{Crosby}},
  \bibnamefont{and} \bibinfo{author}{\bibfnamefont{H.~J.}
  \bibnamefont{Jensen}}, \bibinfo{journal}{Space Science Reviews}
  \textbf{\bibinfo{volume}{198}}, \bibinfo{pages}{3} (\bibinfo{year}{2016}).

\bibitem[{\citenamefont{Dhar}(1990)}]{Dhar_PRL1990}
\bibinfo{author}{\bibfnamefont{D.}~\bibnamefont{Dhar}}, \bibinfo{journal}{Phys.
  Rev. Lett.} \textbf{\bibinfo{volume}{64}}, \bibinfo{pages}{1613}
  (\bibinfo{year}{1990}),
  \urlprefix\url{https://link.aps.org/doi/10.1103/PhysRevLett.64.1613}.

\bibitem[{\citenamefont{Dhar}(1999)}]{Dhar_PhysicaA1999}
\bibinfo{author}{\bibfnamefont{D.}~\bibnamefont{Dhar}},
  \bibinfo{journal}{Physica A: Statistical Mechanics and its Applications}
  \textbf{\bibinfo{volume}{263}}, \bibinfo{pages}{4 } (\bibinfo{year}{1999}),
  ISSN \bibinfo{issn}{0378-4371}, \bibinfo{note}{proceedings of the 20th IUPAP
  International Conference on Statistical Physics},
  \urlprefix\url{http://www.sciencedirect.com/science/article/pii/S0378437198004932}.

\bibitem[{\citenamefont{Vespignani et~al.}(1998)\citenamefont{Vespignani,
  Dickman, Mu\~noz, and Zapperi}}]{Dickman_PRL1998}
\bibinfo{author}{\bibfnamefont{A.}~\bibnamefont{Vespignani}},
  \bibinfo{author}{\bibfnamefont{R.}~\bibnamefont{Dickman}},
  \bibinfo{author}{\bibfnamefont{M.~A.} \bibnamefont{Mu\~noz}},
  \bibnamefont{and} \bibinfo{author}{\bibfnamefont{S.}~\bibnamefont{Zapperi}},
  \bibinfo{journal}{Phys. Rev. Lett.} \textbf{\bibinfo{volume}{81}},
  \bibinfo{pages}{5676} (\bibinfo{year}{1998}),
  \urlprefix\url{https://link.aps.org/doi/10.1103/PhysRevLett.81.5676}.

\bibitem[{\citenamefont{Manna}(1991)}]{Manna_JPhysA1991}
\bibinfo{author}{\bibfnamefont{S.~S.} \bibnamefont{Manna}},
  \bibinfo{journal}{Journal of Physics A: Mathematical and General}
  \textbf{\bibinfo{volume}{24}}, \bibinfo{pages}{L363} (\bibinfo{year}{1991}),
  \urlprefix\url{https://doi.org/10.1088%2F0305-4470%2F24%2F7%2F009}.

\bibitem[{\citenamefont{Dickman et~al.}(2001)\citenamefont{Dickman, Alava,
  A.~Mu\~noz, Peltola, Vespignani, and Zapperi}}]{Dickman_PRE2001}
\bibinfo{author}{\bibfnamefont{R.}~\bibnamefont{Dickman}},
  \bibinfo{author}{\bibfnamefont{M.}~\bibnamefont{Alava}},
  \bibinfo{author}{\bibfnamefont{M.}~\bibnamefont{A.~Mu\~noz}},
  \bibinfo{author}{\bibfnamefont{J.}~\bibnamefont{Peltola}},
  \bibinfo{author}{\bibfnamefont{A.}~\bibnamefont{Vespignani}},
  \bibnamefont{and} \bibinfo{author}{\bibfnamefont{S.}~\bibnamefont{Zapperi}},
  \bibinfo{journal}{Phys. Rev. E} \textbf{\bibinfo{volume}{64}},
  \bibinfo{pages}{056104} (\bibinfo{year}{2001}),
  \urlprefix\url{https://link.aps.org/doi/10.1103/PhysRevE.64.056104}.

\bibitem[{\citenamefont{Dickman et~al.}(2002)\citenamefont{Dickman, Tom\'e, and
  de~Oliveira}}]{Dickman_PRE2002}
\bibinfo{author}{\bibfnamefont{R.}~\bibnamefont{Dickman}},
  \bibinfo{author}{\bibfnamefont{T.}~\bibnamefont{Tom\'e}}, \bibnamefont{and}
  \bibinfo{author}{\bibfnamefont{M.~J.} \bibnamefont{de~Oliveira}},
  \bibinfo{journal}{Phys. Rev. E} \textbf{\bibinfo{volume}{66}},
  \bibinfo{pages}{016111} (\bibinfo{year}{2002}),
  \urlprefix\url{https://link.aps.org/doi/10.1103/PhysRevE.66.016111}.

\bibitem[{\citenamefont{Marro and Dickman}(2005)}]{Marro_Dickman}
\bibinfo{author}{\bibfnamefont{J.}~\bibnamefont{Marro}} \bibnamefont{and}
  \bibinfo{author}{\bibfnamefont{R.}~\bibnamefont{Dickman}},
  \emph{\bibinfo{title}{Nonequilibrium Phase Transitions in Lattice Models}}
  (\bibinfo{publisher}{Cambridge University Press}, \bibinfo{year}{2005}), ISBN
  \bibinfo{isbn}{9780521019460},
  \urlprefix\url{https://books.google.co.in/books?id=80YF69jbczYC}.

\bibitem[{\citenamefont{Dickman et~al.}(1998)\citenamefont{Dickman, Vespignani,
  and Zapperi}}]{Dickman_PRE1998}
\bibinfo{author}{\bibfnamefont{R.}~\bibnamefont{Dickman}},
  \bibinfo{author}{\bibfnamefont{A.}~\bibnamefont{Vespignani}},
  \bibnamefont{and} \bibinfo{author}{\bibfnamefont{S.}~\bibnamefont{Zapperi}},
  \bibinfo{journal}{Phys. Rev. E} \textbf{\bibinfo{volume}{57}},
  \bibinfo{pages}{5095} (\bibinfo{year}{1998}),
  \urlprefix\url{https://link.aps.org/doi/10.1103/PhysRevE.57.5095}.

\bibitem[{\citenamefont{Hwa and Kardar}(1989)}]{Kardar_PRL1989}
\bibinfo{author}{\bibfnamefont{T.}~\bibnamefont{Hwa}} \bibnamefont{and}
  \bibinfo{author}{\bibfnamefont{M.}~\bibnamefont{Kardar}},
  \bibinfo{journal}{Phys. Rev. Lett.} \textbf{\bibinfo{volume}{62}},
  \bibinfo{pages}{1813} (\bibinfo{year}{1989}),
  \urlprefix\url{https://link.aps.org/doi/10.1103/PhysRevLett.62.1813}.

\bibitem[{\citenamefont{Hwa and Kardar}(1992)}]{Kardar_PRA1992}
\bibinfo{author}{\bibfnamefont{T.}~\bibnamefont{Hwa}} \bibnamefont{and}
  \bibinfo{author}{\bibfnamefont{M.}~\bibnamefont{Kardar}},
  \bibinfo{journal}{Phys. Rev. A} \textbf{\bibinfo{volume}{45}},
  \bibinfo{pages}{7002} (\bibinfo{year}{1992}),
  \urlprefix\url{https://link.aps.org/doi/10.1103/PhysRevA.45.7002}.

\bibitem[{\citenamefont{Bonachela and Mu\~noz}(2008)}]{C-DP_Munoz2008}
\bibinfo{author}{\bibfnamefont{J.~A.} \bibnamefont{Bonachela}}
  \bibnamefont{and} \bibinfo{author}{\bibfnamefont{M.~A.}
  \bibnamefont{Mu\~noz}}, \bibinfo{journal}{Phys. Rev. E}
  \textbf{\bibinfo{volume}{78}}, \bibinfo{pages}{041102}
  (\bibinfo{year}{2008}),
  \urlprefix\url{https://link.aps.org/doi/10.1103/PhysRevE.78.041102}.

\bibitem[{\citenamefont{Le~Doussal and Wiese}(2015)}]{Ledoussal_PRL2015}
\bibinfo{author}{\bibfnamefont{P.}~\bibnamefont{Le~Doussal}} \bibnamefont{and}
  \bibinfo{author}{\bibfnamefont{K.~J.} \bibnamefont{Wiese}},
  \bibinfo{journal}{Phys. Rev. Lett.} \textbf{\bibinfo{volume}{114}},
  \bibinfo{pages}{110601} (\bibinfo{year}{2015}),
  \urlprefix\url{https://link.aps.org/doi/10.1103/PhysRevLett.114.110601}.

\bibitem[{\citenamefont{Basu et~al.}(2012)\citenamefont{Basu, Basu,
  Bondyopadhyay, Mohanty, and Hinrichsen}}]{Mohanty_PRL2012}
\bibinfo{author}{\bibfnamefont{M.}~\bibnamefont{Basu}},
  \bibinfo{author}{\bibfnamefont{U.}~\bibnamefont{Basu}},
  \bibinfo{author}{\bibfnamefont{S.}~\bibnamefont{Bondyopadhyay}},
  \bibinfo{author}{\bibfnamefont{P.~K.} \bibnamefont{Mohanty}},
  \bibnamefont{and}
  \bibinfo{author}{\bibfnamefont{H.}~\bibnamefont{Hinrichsen}},
  \bibinfo{journal}{Phys. Rev. Lett.} \textbf{\bibinfo{volume}{109}},
  \bibinfo{pages}{015702} (\bibinfo{year}{2012}),
  \urlprefix\url{https://link.aps.org/doi/10.1103/PhysRevLett.109.015702}.

\bibitem[{\citenamefont{Dickman and da~Cunha}(2015)}]{Dickman_PRE2015}
\bibinfo{author}{\bibfnamefont{R.}~\bibnamefont{Dickman}} \bibnamefont{and}
  \bibinfo{author}{\bibfnamefont{S.~D.} \bibnamefont{da~Cunha}},
  \bibinfo{journal}{Phys. Rev. E} \textbf{\bibinfo{volume}{92}},
  \bibinfo{pages}{020104} (\bibinfo{year}{2015}),
  \urlprefix\url{https://link.aps.org/doi/10.1103/PhysRevE.92.020104}.

\bibitem[{\citenamefont{Grassberger et~al.}(2016)\citenamefont{Grassberger,
  Dhar, and Mohanty}}]{Grassberger_PRE2016}
\bibinfo{author}{\bibfnamefont{P.}~\bibnamefont{Grassberger}},
  \bibinfo{author}{\bibfnamefont{D.}~\bibnamefont{Dhar}}, \bibnamefont{and}
  \bibinfo{author}{\bibfnamefont{P.~K.} \bibnamefont{Mohanty}},
  \bibinfo{journal}{Phys. Rev. E} \textbf{\bibinfo{volume}{94}},
  \bibinfo{pages}{042314} (\bibinfo{year}{2016}),
  \urlprefix\url{https://link.aps.org/doi/10.1103/PhysRevE.94.042314}.

\bibitem[{\citenamefont{Dhar and Pradhan}(2004)}]{Pradhan_JSTAT2004}
\bibinfo{author}{\bibfnamefont{D.}~\bibnamefont{Dhar}} \bibnamefont{and}
  \bibinfo{author}{\bibfnamefont{P.}~\bibnamefont{Pradhan}},
  \bibinfo{journal}{Journal of Statistical Mechanics: Theory and Experiment}
  \textbf{\bibinfo{volume}{2004}}, \bibinfo{pages}{P05002}
  (\bibinfo{year}{2004}),
  \urlprefix\url{https://doi.org/10.1088%2F1742-5468%2F2004%2F05%2Fp05002}.

\bibitem[{\citenamefont{Pradhan and Dhar}(2006)}]{Pradhan_PRE2006}
\bibinfo{author}{\bibfnamefont{P.}~\bibnamefont{Pradhan}} \bibnamefont{and}
  \bibinfo{author}{\bibfnamefont{D.}~\bibnamefont{Dhar}},
  \bibinfo{journal}{Phys. Rev. E} \textbf{\bibinfo{volume}{73}},
  \bibinfo{pages}{021303} (\bibinfo{year}{2006}),
  \urlprefix\url{https://link.aps.org/doi/10.1103/PhysRevE.73.021303}.

\bibitem[{\citenamefont{da~Cunha et~al.}(2009)\citenamefont{da~Cunha, Vidigal,
  da~Silva, and Dickman}}]{Dickman_EPJB2009}
\bibinfo{author}{\bibfnamefont{S.~D.} \bibnamefont{da~Cunha}},
  \bibinfo{author}{\bibfnamefont{R.~R.} \bibnamefont{Vidigal}},
  \bibinfo{author}{\bibfnamefont{L.~R.} \bibnamefont{da~Silva}},
  \bibnamefont{and} \bibinfo{author}{\bibfnamefont{R.}~\bibnamefont{Dickman}},
  \bibinfo{journal}{The European Physical Journal B}
  \textbf{\bibinfo{volume}{72}}, \bibinfo{pages}{441} (\bibinfo{year}{2009}),
  ISSN \bibinfo{issn}{1434-6036},
  \urlprefix\url{https://doi.org/10.1140/epjb/e2009-00367-0}.

\bibitem[{\citenamefont{da~Cunha et~al.}(2014)\citenamefont{da~Cunha, da~Silva,
  Viswanathan, and Dickman}}]{Dickman_JSTAT2014}
\bibinfo{author}{\bibfnamefont{S.~D.} \bibnamefont{da~Cunha}},
  \bibinfo{author}{\bibfnamefont{L.~R.} \bibnamefont{da~Silva}},
  \bibinfo{author}{\bibfnamefont{G.~M.} \bibnamefont{Viswanathan}},
  \bibnamefont{and} \bibinfo{author}{\bibfnamefont{R.}~\bibnamefont{Dickman}},
  \bibinfo{journal}{Journal of Statistical Mechanics: Theory and Experiment}
  \textbf{\bibinfo{volume}{2014}}, \bibinfo{pages}{P08003}
  (\bibinfo{year}{2014}),
  \urlprefix\url{https://doi.org/10.1088%2F1742-5468%2F2014%2F08%2Fp08003}.

\bibitem[{\citenamefont{Jensen et~al.}(1989)\citenamefont{Jensen, Christensen,
  and Fogedby}}]{Jensen_PRB1989}
\bibinfo{author}{\bibfnamefont{H.~J.} \bibnamefont{Jensen}},
  \bibinfo{author}{\bibfnamefont{K.}~\bibnamefont{Christensen}},
  \bibnamefont{and} \bibinfo{author}{\bibfnamefont{H.~C.}
  \bibnamefont{Fogedby}}, \bibinfo{journal}{Phys. Rev. B}
  \textbf{\bibinfo{volume}{40}}, \bibinfo{pages}{7425} (\bibinfo{year}{1989}),
  \urlprefix\url{https://link.aps.org/doi/10.1103/PhysRevB.40.7425}.

\bibitem[{\citenamefont{Kertesz and Kiss}(1990)}]{Kertesz-Kiss1990}
\bibinfo{author}{\bibfnamefont{J.}~\bibnamefont{Kertesz}} \bibnamefont{and}
  \bibinfo{author}{\bibfnamefont{L.~B.} \bibnamefont{Kiss}},
  \bibinfo{journal}{Journal of Physics A: Mathematical and General}
  \textbf{\bibinfo{volume}{23}}, \bibinfo{pages}{L433} (\bibinfo{year}{1990}),
  \urlprefix\url{https://doi.org/10.1088%2F0305-4470%2F23%2F9%2F006}.

\bibitem[{\citenamefont{Manna and Kertész}(1991)}]{Manna-Kertesz1991}
\bibinfo{author}{\bibfnamefont{S.}~\bibnamefont{Manna}} \bibnamefont{and}
  \bibinfo{author}{\bibfnamefont{J.}~\bibnamefont{Kertész}},
  \bibinfo{journal}{Physica A: Statistical Mechanics and its Applications}
  \textbf{\bibinfo{volume}{173}}, \bibinfo{pages}{49 } (\bibinfo{year}{1991}),
  ISSN \bibinfo{issn}{0378-4371},
  \urlprefix\url{http://www.sciencedirect.com/science/article/pii/037843719190250G}.

\bibitem[{\citenamefont{Bak and Paczuski}(1995)}]{Bak-Paczuski1995}
\bibinfo{author}{\bibfnamefont{P.}~\bibnamefont{Bak}} \bibnamefont{and}
  \bibinfo{author}{\bibfnamefont{M.}~\bibnamefont{Paczuski}},
  \bibinfo{journal}{Proceedings of the National Academy of Sciences}
  \textbf{\bibinfo{volume}{92}}, \bibinfo{pages}{6689} (\bibinfo{year}{1995}),
  ISSN \bibinfo{issn}{0027-8424},
  \eprint{https://www.pnas.org/content/92/15/6689.full.pdf},
  \urlprefix\url{https://www.pnas.org/content/92/15/6689}.

\bibitem[{\citenamefont{Yadav et~al.}(2012)\citenamefont{Yadav, Ramaswamy, and
  Dhar}}]{Dhar-PRE2012}
\bibinfo{author}{\bibfnamefont{A.~C.} \bibnamefont{Yadav}},
  \bibinfo{author}{\bibfnamefont{R.}~\bibnamefont{Ramaswamy}},
  \bibnamefont{and} \bibinfo{author}{\bibfnamefont{D.}~\bibnamefont{Dhar}},
  \bibinfo{journal}{Phys. Rev. E} \textbf{\bibinfo{volume}{85}},
  \bibinfo{pages}{061114} (\bibinfo{year}{2012}),
  \urlprefix\url{https://link.aps.org/doi/10.1103/PhysRevE.85.061114}.

\bibitem[{\citenamefont{Hexner and Levine}(2015)}]{Levine-PRL2015}
\bibinfo{author}{\bibfnamefont{D.}~\bibnamefont{Hexner}} \bibnamefont{and}
  \bibinfo{author}{\bibfnamefont{D.}~\bibnamefont{Levine}},
  \bibinfo{journal}{Phys. Rev. Lett.} \textbf{\bibinfo{volume}{114}},
  \bibinfo{pages}{110602} (\bibinfo{year}{2015}),
  \urlprefix\url{https://link.aps.org/doi/10.1103/PhysRevLett.114.110602}.

\bibitem[{\citenamefont{Torquato}(2016)}]{Torquato-PRE2016}
\bibinfo{author}{\bibfnamefont{S.}~\bibnamefont{Torquato}},
  \bibinfo{journal}{Phys. Rev. E} \textbf{\bibinfo{volume}{94}},
  \bibinfo{pages}{022122} (\bibinfo{year}{2016}),
  \urlprefix\url{https://link.aps.org/doi/10.1103/PhysRevE.94.022122}.

\bibitem[{\citenamefont{{Dandekar, Rahul}}(2020)}]{Dandekar-EPL2020}
\bibinfo{author}{\bibnamefont{{Dandekar, Rahul}}}, \bibinfo{journal}{EPL}
  \textbf{\bibinfo{volume}{132}}, \bibinfo{pages}{10008}
  (\bibinfo{year}{2020}),
  \urlprefix\url{https://doi.org/10.1209/0295-5075/132/10008}.

\bibitem[{\citenamefont{Chatterjee et~al.}(2018)\citenamefont{Chatterjee, Das,
  and Pradhan}}]{Chatterjee_PRE2018}
\bibinfo{author}{\bibfnamefont{S.}~\bibnamefont{Chatterjee}},
  \bibinfo{author}{\bibfnamefont{A.}~\bibnamefont{Das}}, \bibnamefont{and}
  \bibinfo{author}{\bibfnamefont{P.}~\bibnamefont{Pradhan}},
  \bibinfo{journal}{Phys. Rev. E} \textbf{\bibinfo{volume}{97}},
  \bibinfo{pages}{062142} (\bibinfo{year}{2018}),
  \urlprefix\url{https://link.aps.org/doi/10.1103/PhysRevE.97.062142}.

\bibitem[{\citenamefont{Eyink et~al.}(1990)\citenamefont{Eyink, Lebowitz, and
  Spohn}}]{Eyink-Lebowitz-Spohn1990}
\bibinfo{author}{\bibfnamefont{G.}~\bibnamefont{Eyink}},
  \bibinfo{author}{\bibfnamefont{J.~L.} \bibnamefont{Lebowitz}},
  \bibnamefont{and} \bibinfo{author}{\bibfnamefont{H.}~\bibnamefont{Spohn}},
  \bibinfo{journal}{Communications in Mathematical Physics}
  \textbf{\bibinfo{volume}{132}}, \bibinfo{pages}{253} (\bibinfo{year}{1990}),
  ISSN \bibinfo{issn}{1432-0916},
  \urlprefix\url{https://doi.org/10.1007/BF02278011}.

\bibitem[{\citenamefont{Kipnis and Landim}(1999)}]{Landim}
\bibinfo{author}{\bibfnamefont{C.}~\bibnamefont{Kipnis}} \bibnamefont{and}
  \bibinfo{author}{\bibfnamefont{C.}~\bibnamefont{Landim}},
  \emph{\bibinfo{title}{Scaling Limits of Interacting Particle Systems}}
  (\bibinfo{publisher}{Springer, Berlin}, \bibinfo{year}{1999}), ISBN
  \bibinfo{isbn}{978-3-662-03752-2},
  \urlprefix\url{https://doi.org/10.1007/978-3-662-03752-2}.

\bibitem[{\citenamefont{Carlson et~al.}(1990)\citenamefont{Carlson, Chayes,
  Grannan, and Swindle}}]{Carlson_PRL1990}
\bibinfo{author}{\bibfnamefont{J.~M.} \bibnamefont{Carlson}},
  \bibinfo{author}{\bibfnamefont{J.~T.} \bibnamefont{Chayes}},
  \bibinfo{author}{\bibfnamefont{E.~R.} \bibnamefont{Grannan}},
  \bibnamefont{and} \bibinfo{author}{\bibfnamefont{G.~H.}
  \bibnamefont{Swindle}}, \bibinfo{journal}{Phys. Rev. Lett.}
  \textbf{\bibinfo{volume}{65}}, \bibinfo{pages}{2547} (\bibinfo{year}{1990}),
  \urlprefix\url{https://link.aps.org/doi/10.1103/PhysRevLett.65.2547}.

\bibitem[{\citenamefont{Eyink et~al.}(1991)\citenamefont{Eyink, Lebowitz, and
  Spohn}}]{Eyink-Lebowitz-Spohn1991}
\bibinfo{author}{\bibfnamefont{G.}~\bibnamefont{Eyink}},
  \bibinfo{author}{\bibfnamefont{J.~L.} \bibnamefont{Lebowitz}},
  \bibnamefont{and} \bibinfo{author}{\bibfnamefont{H.}~\bibnamefont{Spohn}},
  \bibinfo{journal}{Communications in Mathematical Physics}
  \textbf{\bibinfo{volume}{140}}, \bibinfo{pages}{119} (\bibinfo{year}{1991}),
  ISSN \bibinfo{issn}{1432-0916},
  \urlprefix\url{https://doi.org/10.1007/BF02099293}.

\bibitem[{qua(Quasi-periodic)}]{quasi-periodic}
 (\bibinfo{year}{Quasi-periodic}), \bibinfo{note}{to generate a quasi-periodic
  uniform critical density profile with density $\rho<1$, we use the floor
  function to obtain particle number $n_{X} =\lfloor{X \rho} \rfloor -
  \lfloor{(X-1) \rho}\rfloor$ at $X$th site (i.e., $n_X$ is the $X$th bit in
  the generated sequence), where $\lfloor x \rfloor$ is the largest integer
  less than the number $x$; note that $n_X$ takes value $0$ or $1$.}

\end{thebibliography}

\end{document}